\documentclass[usenatbib]{mnras}
\usepackage{natbib,amssymb,amsmath,times,graphicx,url,aas_macros}
\usepackage{setspace}      
\usepackage{epstopdf}       
\usepackage{verbatim}       
\usepackage{bm}		  
\usepackage{fleqn}		  
\usepackage{enumitem}
\usepackage{Wurster_macros}  

\defcitealias{WursterBatePrice2019}{WBP2019}
\defcitealias{WursterRowan2023b}{Paper II}
\newcommand{\tfinal}{1.45t$_\text{ff}$}
\newcommand{\ithree}{\textit{I03}}
\newcommand{\ifive}{\textit{I05}}
\newcommand{\iten}{\textit{I10}}
\newcommand{\itwenty}{\textit{I20}}
\newcommand{\nthree}{\textit{N03}}
\newcommand{\nfive}{\textit{N05}}
\newcommand{\nten}{\textit{N10}}
\newcommand{\ntwenty}{\textit{N20}}
\newcommand{\hyd}{\textit{Hyd}}

\title[Star-forming environments I]{Star-forming environments in smoothed particle magnetohydrodynamics simulations I: Clump extraction and properties}

\author[Wurster \& Rowan]{James Wurster$^{1}$\thanks{jhw5@st-andrews.ac.uk} and Connar Rowan$^{2,1}$\\
$^{1}$Scottish Universities Physics Alliance (SUPA), School of Physics and Astronomy, University of St. Andrews, North Haugh, St Andrews, Fife, KY16 9SS, UK \\
$^{2}$Rudolf Peierls Centre for Theoretical Physics, Clarendon Laboratory, Parks Road, Oxford, OX1 3PU, UK \\
}
\date{Submitted: Revised: Accepted: }
\pagerange{\pageref{firstpage}--\pageref{lastpage}} \pubyear{2023}

\begin{document}
\label{firstpage}
\bibliographystyle{mnras}
\maketitle

\begin{abstract}
What is the nature of a star forming clump?  Observations reveal these to be chaotic environments being modified and influenced by many physical processes.  However, numerical simulations often define these initial star forming clumps to be idealised objects.  
In this paper, we define and analyse 109 star forming clumps extracted from our previous low-mass star cluster simulations.  To define a clump, we identify all the gas in a simulation that ever becomes bound to or accreted onto a star, then follow the gas backwards in time until it decreases to a critical density.  This gas, and its neighbouring gas, is defined as our star forming clump.  Our clumps span a mass range of $0.15 \lesssim M/$M$_\odot \lesssim 10.2$, while the density range within each clump spans 2--4 orders of magnitude.  The gas density distribution is not smooth, indicating that it is highly structured.  The clumps are turbulent, with no coherent rotation.  Independent of the initial magnetic field strength of the parent cloud, all clumps yield a similar range of field strengths.  The clump magnetic field is ordered, but not reflective of the initial field geometry of the parent cloud.  In general, most clump properties have a slight trend with clump mass but are independent of (or only very weakly dependent on) the properties of the parent cloud.  We conclude that stars are born from a wide variety of environments and there is not a single universal star forming clump. 
\end{abstract}


\section{Introduction}
\label{intro}

Star forming regions are messy and chaotic environments, with structures on many scales.    While the entire region is typically referred to as a \textit{cloud}, dense regions embedded within the cloud are \textit{clumps}, while the very dense regions in the clumps are \textit{cores} \citep{BerginTafalla2007}; there is general agreement of the three terms, although there is ambiguity amongst the specific definitions and divisions between the levels.  Cores are the smallest of the three levels, and are the immediate precursor to star formation.  These regions are often small and somewhat spherical \citepeg{MyersLinkeBenson1983,Myers+1991,JijinaMyersAdams1999}, and often approximated as Bonnor-Ebert spheres \citep{Ebert1955,Bonnor1956}.  There has been considerable effort measuring the masses of these cores to define a Core Mass Function (CMF) and to relate it to the Initial Mass Function (IMF), both numerically and observationally \citepeg{Klessen2001,TilleyPudritz2004,TilleyPudritz2007,AlvesLombardiLada2007,NutterWardthompson2007,Padoan+2007,ChabrierHennebelle2010,Schmidt+2010,Marsh+2016,Sokol+2019,NtormousiHennebelle2019,Konyves+2020,Ladjelate+2020,Pelkonen+2021,Takemura+2021,Pouteau+2022,Padoan+2023}.  

To understand star formation, is it enough to simply understand the properties of the core (technically a pre-stellar core since the star itself has yet to form)?  The core collapse model of \citet{MckeeTan2002,MckeeTan2003} suggests that this core itself consists of the entire mass reservoir required to form the star.  The competitive accretion model of \citet{Zinnecker1982} and \citet{Bonnell+2001ca,Bonnell+2001acc} suggests that a significant fraction of the stellar mass comes from outside of the core, and is accreted through gravitational processes.  The inertial-inflow model of \citet{Padoan+2020} also suggests that a significant fraction of the stellar mass comes from outside of the core, however, the infall is through turbulent-generated structures rather than gravity.  The latter two models clearly suggest that knowledge of the core is not enough to predict its evolution and possible collapse into a star.  

Nonetheless, there have been numerous numerical studies to investigate star formation originating from a core, using either an initial core of uniform density or centrally condensed as an (e.g.) supercritical Bonnor-Ebert sphere.  Given the idealised initial core of (typically) low-mass, high resolution simulations have been performed to provide invaluable information on star formation, the resulting objects, the physical and numerical processes.  For example, simulations initialised from idealised initial conditions have greatly expanded our knowledge on 
radiation hydrodynamics \citepeg{Tomida+2010rmhd,Tomida+2013,Bate2011}, 
ideal magnetic fields \citepeg{BateTriccoPrice2014}, 
non-ideal magnetic fields \citepeg{Tsukamoto+2015hall,Tsukamoto+2015oa,Tsukamoto+2017,WursterPriceBate2016,WursterBatePrice2018sd,WursterBatePrice2018hd,WursterBatePrice2018ff,WursterBatePrice2019,WursterBateBonnell2021,Wurster+2022,MarchandCommerconChabrier2018,Vaytet+2018}, 
magneto-turbulence \citepeg{TsukamotoMachida2013,Joos+2013,WursterLewis2020d,WursterLewis2020sc}, 
angular momentum transport \citepeg{WursterBatePrice2018hd,WursterBateBonnell2021,MisugiInutsukaArzoumanian2019,Marchand+2019,Marchand+2020},  
disc formation \citepeg{MachidaInutsukaMatsumoto2010,MachidaMatsumoto2011,MachidaMatsumotoInutsuka2016,InutsukaMachidaMatsumoto2010,WursterPriceBate2016,WursterBatePrice2018hd,WursterBatePrice2019,WursterBate2019,WursterBateBonnell2021,WursterLewis2020d,Tomida+2017}, 
outflows \citepeg{MachidaInutsukaMatsumoto2008,Machida2014,Machida2021,WursterBatePrice2018sd,WursterBateBonnell2021,Wurster+2022,HiguchiMachidaSusa2019}, 
dust evolution \citepeg{BateLorenaguilar2017,TsukamotoMachidaInutsuka2021sf,TsukamotoMachidaInutsuka2021ash,Lebreuilly+2023,Bate2022}, 
artificial resistivity \citepeg{Wurster+2017}, 
and sink particles \citepeg{MachidaInutsukaMatsumoto2014}.  

One important conclusion from studying isolated star formation initialised from idealised initial conditions concerns the magnetic braking catastrophe: rotationally supported discs did not form in simulations that included strong, ideal magnetic fields \citepeg{AllenLiShu2003,MellonLi2008}.  Several solutions to this catastrophe were suggested with the robust solution being the inclusion of the non-ideal magnetohydrodynamic (MHD) term of the Hall effect \citepeg{Tsukamoto+2015hall,WursterPriceBate2016,WursterBateBonnell2021}; this solution held even when the initial cores included subsonic turbulence \citet{WursterLewis2020d,WursterLewis2020sc}.  However, when larger, more massive, and more chaotic systems were simulated, it became clear that the magnetic braking catastrophe was a result of the idealised initial conditions \citepeg{Seifried+2013,WursterBatePrice2019}.

Despite the above advances, and in part due to the solution of the magnetic braking catastrophe, how realistic are isolated star formation simulations given that a star forming core is not an isolated object (assuming the competitive accretion or inertial inflow models best describes star formation)?  To better understand star formation and the resulting star that will form from a core, knowledge of its progenitor clump is required.  This is a necessary yet ambiguous step since `clump' is not a well-defined object.  

There have been several simulations that attempt to define a progenitor clump, or at least determine where the gas comes from that is ultimately accreted onto a star \citepeg{LewisBatePrice2015,LewisBate2017,Pelkonen+2021,ArroyocyhavezVazquezsemadeni2022,CollinsLeJimenezvela2023}.  

\citet{ArroyocyhavezVazquezsemadeni2022} identified clumps at a given time, then they traced those particles backwards in time for a set length of time.  Moving backwards in time, the clumps expanded from \sm1 to several pc, and the density decreased by \sm2 orders of magnitude; when they constructed a box around the selected particles, up to 75 per cent of the particles in the box did not ultimately end up in their clump.

The goal of \citet{Pelkonen+2021} was to investigate the CMF, thus they focused on the cores just prior to the birth of a star.  Most of their cores had masses that were less than the final mass of the sink particles that they spawned, indicating that the stars accreted material from sources other than just their progenitor core.  To determine the extent from which stars accreted their material, they identified all the tracer particles that ended in a given star and traced them back to the beginning of the simulation\footnote{This is a turbulence-in-a-box simulation, and `beginning' is define to occur once the stirring is turned off and gravity is turned on.}.  The spheres containing 95 per cent of these particles ranged in size from a few to several thousand au, which was \sm1-1000 times larger than the radius of their defined cores.  These conclusions, and the conclusions form other numerical studies, suggest that even once a core is formed, knowledge of it is not enough to predict how it will evolve or the properties of the resulting star.

In this study, we use the low-mass star cluster simulations of \citet[][herein \citetalias{WursterBatePrice2019}]{WursterBatePrice2019} to extract regions of gas that will ultimately form or influence a star.  Given that these will be extended regions, we will refer to them as \textit{clumps} rather that cores; this means that we will not be able to comment on the CMF.  We will set criteria for how far back in time we trace the gas rather than pre-selecting a length of time or a fixed time in the simulation.  We also construct non-regular boundaries, rather than a sphere or a box.  The simulation in \citetalias{WursterBatePrice2019} used the smoothed particle hydrodynamics (SPH) method, thus we track the motion of the exact Lagrangian particles that influence star formation and evolution rather than requiring tracer particles; if not treated carefully, errors can compound on tracer particles so that they cease to accurately trace the flow \citep{Pelkonen+2021}.  Our low-mass star clusters are much less massive than the studies of (e.g.) \citet{Pelkonen+2021} and \citet{ArroyocyhavezVazquezsemadeni2022}, thus we naturally yield smaller clumps but at much higher numerical resolution.

Our ultimate goal is to construct initial star-forming clumps that are low-mass as in simulations of isolated star formation but structured as in cluster simulations.  These clumps can then themselves be studied (as below) and used as initial conditions for future simulations of isolated star formation.  The latter is performed in our companion paper, Wurster \& Rowan (submitted; herein \citetalias{WursterRowan2023b}), where we evolve several of these clumps to determine how closely their evolution agrees with the gas in the cluster simulations themselves.  Our long-term aim is to repeat the study of \citetalias{WursterRowan2023b} but at the numerical resolutions similar to current simulations in the literature of isolated star formation in order to perform a detailed investigation into star formation and all the related processes in a realistic environment at a high resolution.

The rest of the paper is organised as follows.  In \secref{sec:cluster} we  summarise the star cluster simulations of \citet{WursterBatePrice2019}.  In \secref{sec:ext} we discuss our extraction method by which we generate our initial conditions.  In \secref{sec:results} we analyse the extracted initial star forming clumps, and discuss the caveats in \secref{sec:disc}.  We conclude in \secref{sec:conc}.

\section{Low-mass star clusters}
\label{sec:cluster}
This study uses the data from \citetalias{WursterBatePrice2019}, which we define as the \textit{parent simulation} or \textit{parent cloud}.  We will briefly summarise the relevant information from that paper, but we refer the reader to that paper for the details.  

In that study, using the equations of self-gravitating radiation magnetohydrodynamics, we modelled the evolution of a 50~\Msun{} cloud using the radiation smoothed particle non-ideal magnetohydrodynamics code \textsc{sphNG} \citep{Benz1990}.  The initial cloud was a sphere of radius $r = 0.1875$~pc and uniform density \rhotwoeq{1.22}{-19}, and was embedded in a warm medium that had a density 30 times lower.  The cloud had a temperature of 8.8~K at the centre, which increased to 13~K at the edge. The cloud was seeded with an initial turbulent velocity field following the method of \citet{OstrikerStoneGammie2001} and \citet{BateBonnellBromm2003}; the rms Mach number was $\mathcal{M} = 4.8$, which was set so that the initial kinetic energy was equal to the initial potential energy\footnote{Note that the initial Mach number was incorrectly given in \citetalias{WursterBatePrice2019}; that value was copied from \citet{BateBonnellBromm2003} where the initial cloud was isothermal at 10~K and did not include an initial temperature gradient as in \citetalias{WursterBatePrice2019}.}.  There was no bulk rotation.  Finally, the cloud and medium where threaded with a uniform magnetic field of $\bm{B} = -B_0\hat{\bm{z}}$.  

Sink particles \citep{BateBonnellPrice1995} with an accretion radius of $r_\text{acc} = 0.5$~au were employed to represent individual stars; sinks were permitted to merge if they came within 27~\Rsun{} of one another.  We used $5\times10^6$ equal mass SPH particles in the sphere for a mass resolution of $m_\text{particle} = 10^{-5}$~\Msun{} per particle.
We used the $M_4$ cubic spline kernel, thus the properties of particle $a$ were calculated using all neighbours $b$ within 2$h$, where $h$ is the smoothing length.

The aim of that study was to investigate the effect of the magnetic field strength and of ideal and non-ideal magnetic fields on the evolution of the a low-mass cluster.   \citetalias{WursterBatePrice2019} included nine models as summarised in Table~\ref{tab:models}, and we use the same model names here.  The models were evolved for at least 1.45~\tff{} = 276~kyr, where \tff{} is the free-fall time, and in this study, we use only the data up to this time; the final column in Table~\ref{tab:models} list the number of sink particles in each cloud at 1.45~\tff{}.  Sink masses ranged from 0.01 - 3~\Msun{}, however, the small number of sinks prevented a determination and analysis of the IMF or an investigation into the CMF.  The majority of the stars formed multiple systems, where we tracked systems up to order four \citep[as in][]{Bate2018}.  

All models formed discs, including discs around single and multiple stars, and sometimes circumsystem discs where the component stars had their own circumstellar discs.  Given the ubiquity of discs, even in models with initially strong ideal magnetic field strengths, we concluded that there was no magnetic braking catastrophe.  In agreement with \citet{Seifried+2013}, we concluded that the magnetic breaking catastrophe that appeared in simulations of isolated star formation \citepeg{AllenLiShu2003,PriceBate2007,MellonLi2008,HennebelleFromang2008,LiKrasnopolskyShang2011} due to unrealistic initial conditions.  This is hence the motivation for this study, where we aim to understand realistic initial star forming conditions.  Although our discs were resolved in \citetalias{WursterBatePrice2019}, they were still at low resolution so only crude bulk properties could be determined.  Moreover, there were no first core outflows, however, in \citet{Wurster+2022}, we confirmed that this was a result of resolution and not environment.

\begin{table}
\begin{center}
\begin{tabular}{c c c c c }
\hline
Name              &    $\mu_0$ &  $B_0$ [$10^{-5}$G] & $N_\text{stars}$  & $N_\text{mergers}$ \\
\hline 
\emph{N03}    & 3 & 6.48 &  10 & 0  \\ 
\emph{N05}    & 5 & 3.89 & 17 &  0 \\ 
\emph{N10}    & 10 & 1.94 &   8 &  0 \\ 
\emph{N20}    & 20 & 0.972 & 10 & 1   \\ 
\emph{Hyd}    & $\infty$ & 0 & 19 & 0     \\ 
\emph{I03}     & 3 & 6.48 &  11 & 1   \\ 
\emph{I05}     & 5 & 3.89 & 11 & 1 \\ 
\emph{I10}     & 10 & 1.94 &   7 & 0 \\ 
\emph{I20}     & 20 & 0.972 &11 & 2          \\ 
\hline
\end{tabular}
\caption{Summary of the models from \citetalias{WursterBatePrice2019}.  The model names are as in that paper, where models starting with \textit{N} employ non-ideal MHD and models starting with \textit{I} employ ideal MHD.  The second and third columns give the simulations' initial magnetic field strength in terms of the critical mass-to-flux ratio and in physical units, respectively.  The fourth column gives the number of stars (sinks) at 1.45\tff{}, and the fifth column gives the number of mergers of star particles; the number of initial clumps in this study is the sum of the final two columns. }
\label{tab:models} 
\end{center}
\end{table}

\section{Generating initial star forming clumps}
\label{sec:ext}

Our first goal is to generate an initial star forming clump using the data in \citetalias{WursterBatePrice2019}.  We will define a clump to be comprised of all the gas that is ultimately accreted onto a star (i.e., sink particle), is associated with the star (e.g., in a disc), or is initially casually connected to the aforementioned gas.  
The first step is to extract this gas to define the clump (Sections \ref{sec:ext:ext} and \ref{sec:ext:aug}), and the second step is to add a background medium (\secref{sec:ext:gen}) so that the clumps can be numerically evolved in \citetalias{WursterRowan2023b} and future studies.  
\subsection{Extracting star forming gas}
\label{sec:ext:ext}
To extract the initial clump, we compile a list of SPH particles that are associated with each sink throughout its entire evolution.  While we can track the movement of every particle through its unique ID number, \textsc{sphNG} unfortunately does not record the IDs of particles that are accreted onto sinks; therefore, we need to make some assumptions about which particles are added to the list.  This also generalises our process to other SPH codes.

The steps to create our list is outlined below, and is naturally the same for all sinks;  particles may appear on multiple lists, but given the number of higher order systems that form, this is to be expected.
\begin{enumerate}[label={\arabic*.}]
\item At the final time $t_\text{final}$, identity a sink particle.
\item Add to the list every gas particle that is within $r_\text{min}$.
\item Add to the list every bound gas particle within $r_\text{max}$ that has a density of $\rho > \rho_\text{min}$ and has an eccentricity of $e < e_\text{max}$.
\item Define the final clump mass, $M_\text{clump}^\text{final}$, to be the mass of all the gas particles on the list plus the sink particle mass.
\item Analyse sequential dump files, moving backwards in time.  For each dump, perform the following actions:
   \begin{enumerate}
   \item Repeat steps 2 \& 3, adding new particles to the list as required.
   \item Determine the current clump mass, $M_\text{clump}$, which is the mass of all the particles on the updated list plus the sink's \textit{current} mass.
   \item While $M_\text{clump} < M_\text{clump}^\text{final}$, add gas particles to the list by selecting the nearest particles not already on the list; this is required to ensure we account for all the accreted particles.
   \end{enumerate}
\item Repeat step 5 until a dump file is reached where the sink does not exist.  On this dump, perform the following actions:
\begin{enumerate}
   \item Identify the gas particle that becomes the sink; in \textsc{sphNG}, the progenitor gas particle will have the same ID as the sink particle it becomes.  
   \item Create a `pseudo-sink', which is a group of gas particles that has the same mass as the sink on the next dump  (i.e., on the first dump when the sink exists) and is centred on the progenitor gas particle.  
   \item Calculate the bulk velocity of the pseudo-sink.
   \item Add all of these particles in the pseudo-sink to the list since, until now, they have most likely been part of the sink particle itself. 
\end{enumerate}
\item Repeat step 5 one final time, using the mass and velocity of the pseudo-sink in place of the mass and velocity of an actual sink.  This is the complete list of particles.
\item Track these particles backwards through time until the maximum density drops below a given extraction density, $\rho_\text{ext}$.  The initial conditions for the sink particle is finally defined as the particles on the list in this dump file.
\end{enumerate}

We define this list of particles to comprise the \textit{associated clump}.  We define the time of the dump file in step 8 to be the \textit{extraction time}.

For the data in \citetalias{WursterBatePrice2019}, we define the final time to be $t_\text{final} = $ \tfinal{} = 276~kyr, and set $r_\text{min}$ = 50~au, $r_\text{max}$ = 1000~au, $\rho_\text{min} = 10^{-18}$~\gpercc{}, $e_\text{max} = 0.5$, and $\rho_\text{ext} = 10^{-16}$~\gpercc{}.  The isothermal collapse phase of star formation last until \rhoxapprox{-13} and the evolutionary dependence of the maximum magnetic field strength on density changes at \rhoxapprox{-15} \citep[see fig. 2 of][]{WursterBatePrice2018sd}; therefore, any value of $\rho_\text{ext}$ smaller than these would be reasonable.  We empirically selected the above value to ensure that the clumps were not too large and that they would form at least one star within a reasonable length of time.  Tests show that the results are relatively insensitive to $r_\text{max}$ due to the boundness criteria.

\subsection{Augmenting the star forming gas}
\label{sec:ext:aug}
The particles in the associated clump as identified in  \secref{sec:ext:ext} represent the particles that will ultimately become the star or become associated with (e.g., bound to) it.  We now modify this list for numerical stability and so that the final clump is a true representation of the extracted region in the parent cloud.    

First, we remove particles that are too distant and quasi-isolated from the bulk of the associated list; these particles will likely not reasonably contribute to the evolution of the clump, and if they do, their evolution may be artificially influenced by the background gas that will be added in the next section.  We remove these distant particles by making a new list that initially contains the densest gas particle from the original list and all particles within 250~au.  We then expand this new list by adding all old-list particles within 2$h$ of a new-list member\footnote{If the parent simulation uses a kernel other than the cubic spline, then the value of 2$h$ must be modified to agree with the employed kernel.}; we continue this process iteratively until we have found all old-list particles that are connected to the new list in this manner.  We discard the remaining old-list particles that are not on the new list.  In most clumps, this is a small percentage of particles and are the particles that have come in from very far away.  

In some clumps where the associated gas is spread out, this process can discard the majority of the particles, which will yield clumps that are not representative of the associated gas.  To prevent this, we enforce that at least 90 per cent of the associated gas particles must be retained, where 90 per cent was empirically chosen.  To achieve this, if no particles are added to the new list but this threshold it not reached, we perform a single search and add particles within 4$h$ of any new-list members.  After this search, we revert to 2$h$, but we repeat this process as needed until 90 per cent of the associated gas is on the new list.  See \secref{sec:diffuse} for additional discussion regarding this threshold.

Second, we loop though \textit{all} particles, and add all particles that are within 4$h$ of a particle on our new list.  
Without this step, the gas density in the clump will be lower than its counterpart in the parent cloud since there is no guarantee that all the neighbours of a particle in the associated clump will also be in the associated clump; recall that the density of SPH particle $a$ is calculated using all of its neighbours $b$ \citerev{Price2012}.  We select a value of $4h$ to ensure accurate boundaries, and that the properties of the associated particles near the edge of the cloud are accurately calculated.  
While this adds gas that will not be associated with the final system, it is required to provide the proper initial properties of the gas particles that is on the associated list.  

This new group is our initial star forming clump, which we define as the \textit{augmented clump} or simply \textit{clump}.

\subsection{Generating background medium}
\label{sec:ext:gen}
Given that we evolved magnetic fields in the parent simulations and do so in \citetalias{WursterRowan2023b}, we require boundary conditions.  We use the `sphere-in-box' setup\footnote{The `sphere-in-box' is a cold, dense sphere of gas embedded in a box of warm, low-density gas where the two media are in pressure equilibrium.  The box is periodic and has a size four times the sphere's radius.} of our previous simulations \citepalias[including][]{WursterBatePrice2019} as motivation for the current background.  

We calculate the average magnetic field, density, and thermal energy of the boundary particles, where we identify the boundary particles of the clump to be those with less than 30 neighbours.  Given that we use the $M_4$ cubic-spline kernel, particles should have \sm58 neighbours, and we have confirmed that this is true for the particles away from the boundary.  We then add background particles on a cubic lattice\footnote{This lattice was chosen for ease.} with a density of 30 times lower than the average of the boundary particles.  The background particles are given the magnetic field of the average of the boundary, and a thermal energy 30 times higher than the average so that, on average, we have pressure equilibrium at the boundary.  The background has a dimension of $\{2L_\text{x},2L_\text{y},2L_\text{z}\}$, where $\{L_\text{x},L_\text{y},L_\text{z}\}$ is the dimension of the smallest cuboid we can generate to encompass the clump, with the centre-of-mass of the clump at the centre of the box.   Finally, we remove all background particles that come within $h$ of the clump to minimise mixing.  The entire domain is periodic.  

This background medium is excluded in all analysis in the reminder of this paper.

\subsection{Development notes}
\subsubsection{Multiplicity}
The design of the above algorithm is based upon an assumption that a star forms and lives in isolation.  This is clearly not true, and the majority of the stars in \citetalias{WursterBatePrice2019} are in multiple systems.  However, how to deal with multiplicity is non-trivial.  One option would be to use the stellar system mass at every dump rather than the mass of the single star.  While this would be reasonable for a tight binary, it could lead to issues for wide systems.  A second option would be to use the above method, but merge the associated lists of stars that are bound at any point in time.  Using either of these methods would yield fewer unique clumps since they would be based upon systems and not single stars; fewer cores, however, is not detrimentally to this study and its successors.  Tests have shown that the latter option yields very massive clumps that are a large fraction of the parent cloud, and these massive clumps defeat the goal of this study to create low-mass clumps.

In this study, we chose the above parameters and algorithms as a compromise to identify the gas associated with each single star while promoting the extraction of low-mass clumps.  These clumps will be analysed in \secref{sec:results}, while a future study may investigate how the various choices of parameters and algorithms affect the clumps.  

\subsubsection{Sink mergers}
Sinks are permitted to merge in \citetalias{WursterBatePrice2019}.  When this occurred, sink $A$ was given the centre-of-mass and -momentum of the pair while the sink $B$ was numerically killed.  For sink $B$, we begin the process of creating its clump beginning with the final dump in which it existed as an individual star.  Therefore, step 1 begins at $t < t_\text{final}$.  We then proceed as described.

We take no special actions with respect to sink $A$.  Although the gas in sink $B$ is not explicitly accounted for in the list of sink $A$, we find that nearly all of the augmented gas that resides in clump of sink $B$ also resides in the clump of sink $A$.  Therefore, explicit actions are unnecessary. 

\subsection{Definitions}
\label{sec:ext:def}
For simplicity and clarity, we re-state the important definitions here that we will use throughout the paper.
\begin{itemize}
\item \textit{Parent simulation} or \textit{parent cloud}:  The simulations presented in \citetalias{WursterBatePrice2019}.
\item \textit{Associated clump}: The complete list of SPH particles that will ultimately accrete onto the sink, come within $r_\text{min}$ of the sink, or become bound to the sink; see \secref{sec:ext:ext} for construction.
\item \textit{Augmented clump} or \textit{clump}: The final list of SPH particles that includes the associated particles plus all their neighbours, as constructed in \secref{sec:ext:aug}
\item \textit{Extraction time}: The time at which the maximum density in the associated clump decreases to $\rho < \rho_\text{ext}$, defining the initial location of the associated clump.
\item \textit{Massive clump}: An augmented clump with $M > 5$~\Msun{}.
\item \textit{Diffuse clump}: A clump with a ratio of associated gas mass to total gas mass of $\lesssim2$ per cent.
\end{itemize}

\section{Results}
\label{sec:results}
By analysing every sink that formed in \citetalias{WursterBatePrice2019}, we extracted 109 clumps, of which 19 are from \textit{Hyd} and have no magnetic properties.  There have been five sink mergers, therefore the extraction of five clumps began at $t < t_\text{final}$, and another five clumps accounted for the merger through the generation of the augmented list.  There are 16 clumps that required temporarily expanding the search radius to ensure that 90 per cent of the associated gas was retained for the augmented clump.  These descriptions of special clumps are not mutually exclusive, and, with the exception of the magnetic field, we do not distinguish between them and the remainder of the clumps. 

\begin{figure} 
\centering
\includegraphics[width=\columnwidth]{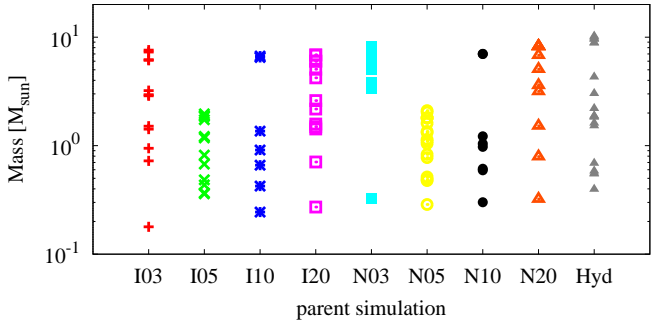}
\caption{The mass of each clump from each parent cloud.  Not all clouds contain clumps that span the full range of $0.15 \lesssim M/$\Msun{}$ \lesssim 10.2$, demonstrating the varied nature of each parent simulation. }
\label{fig:cloudmass}
\end{figure}

\begin{figure} 
\centering
\includegraphics[width=\columnwidth]{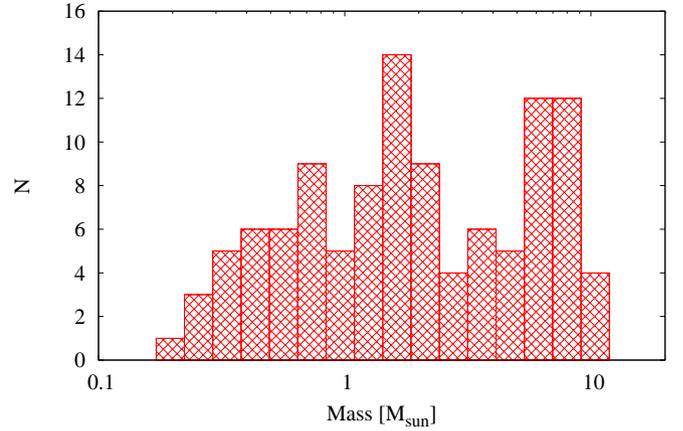}
\caption{Histogram of the clump masses.  The range is larger than the typical range in the literature for the initial cores of isolated star formation models.  The most massive clumps are \sm10-20 per cent of the total mass of the parent cloud.}
\label{fig:masshisto}
\end{figure}

\figref{fig:cloudmass} shows the mass of each clump that was extracted from each parent simulation, and \figref{fig:masshisto} shows a histogram of the (augmented) clump masses.  
There is an approximate Gaussian distribution of clump masses when considering all 109 clumps, although there is a second local maximum at high clump mass. Clumps exist in the range from 0.15 to 10.2~\Msun{}, which is a larger range than than the typical range in the literature of 1-5~\Msun{} for the initial core in isolated star formation models\footnote{This is the common initial mass range as used in the studies listed in the Introduction.}.  Most parent simulations yield clumps that span the majority of the mass range; however, \ifive{} and \nfive{} yield no high-mass clumps, and \nthree{} yields only a single clump below \sm 3~\Msun{}.  \hyd{} yields seven massive clumps that have a high overlap of constituent gas (as per visual inspection); this is a larger number of massive clumps (relatively and absolutely) than in the magnetised simulations, suggesting that magnetic fields regulate the gas flow and that more distant gas is less likely to affect the evolution of a stellar system. 

The average ratio of associated-to-augmented clump mass is $0.18 \pm 0.10$, where the uncertainty represent one standard deviation; see \figref{fig:massratio}.  This is slightly lower than the ratio identified in \citet{ArroyocyhavezVazquezsemadeni2022}.  There is no explicit trend in ratio with augmented clump mass, however, this plot shows that we are required to add a reasonable amount of gas to all clumps to properly map the initial density distribution and account for reasonable boundaries (see \secref{sec:ext:aug}).  There are a few clumps where $\lesssim2$ per cent of the clump mass is from the associated gas, indicating that this gas has travelled a considerable distance; these diffuse clumps tend to be from the hydrodynamic parent simulation or MHD simulations with weak magnetic fields, providing further evidence of how strong magnetic fields constrain the flow of the gas.
\begin{figure} 
\centering
\includegraphics[width=\columnwidth]{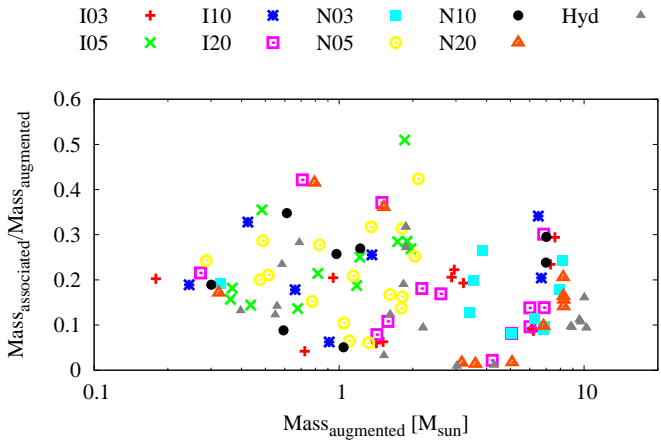}
\caption{Ratio of the associated to augmented clump masses.  There is no trend with mass or initial magnetic field strength of the parent clouds.  The average ratio is $0.18 \pm 0.10$.}
\label{fig:massratio}
\end{figure} 
The distinction between massive and low-mass clumps at $M \approx $ 5~\Msun{} is somewhat arbitrary from this histogram (\figref{fig:masshisto}), but is more evident when excluding the diffuse clumps; see \figref{fig:massratio}.

The final mass of the sink particle used to generate the clump has $0.87 \pm 0.13$ of the mass in the associated clump, which is in reasonable agreement with isolated star formation models \citepeg{WursterPriceBate2016}; note, however, that this ratio decreases to $0.16 \pm 0.09$ when comparing the final sink mass to the augmented clump mass.  Therefore, unsurprisingly, the augmented gas may be required to properly model the evolution of the associated gas, but it does not end up accreting onto the star itself.  We again remind the reader that our extraction process, and hence these ratios, originates with a single sink and does not account for multiplicity, which is prevalent at the end of \citetalias{WursterBatePrice2019}.

\figref{fig:morph} shows four example clumps extracted from \textit{N05} compared to the parent simulation at extraction time.  All of our augmented clumps are shown in \figref{fig:morph:all}. 
\begin{figure*} 
\centering
\includegraphics[width=\textwidth]{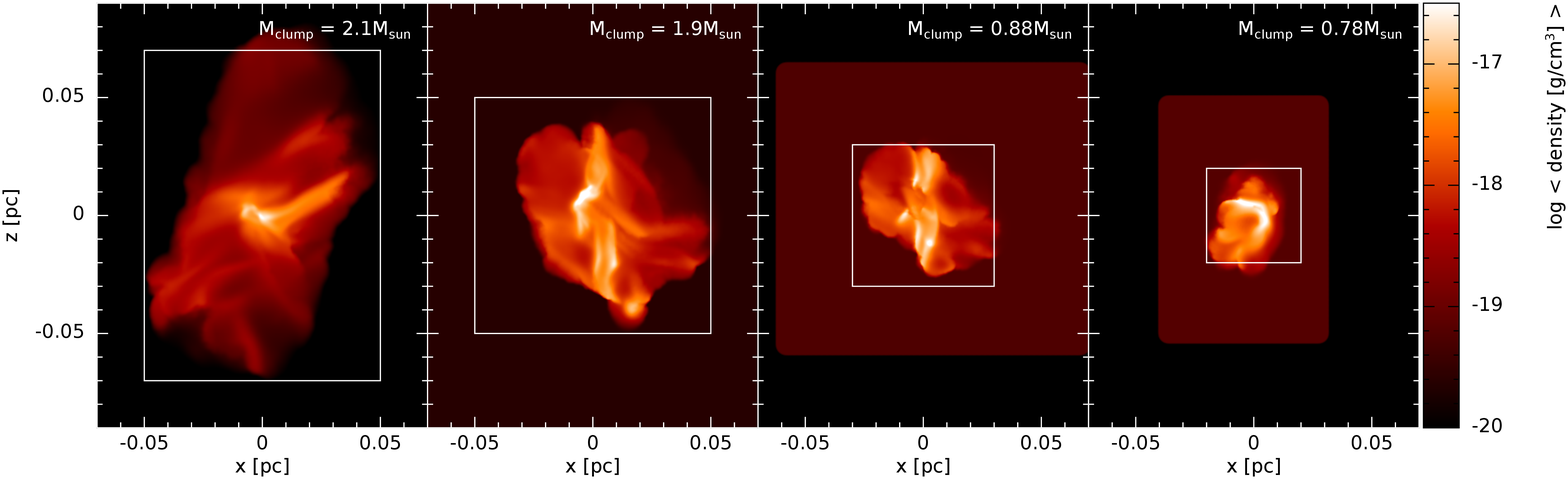}
\includegraphics[width=\textwidth]{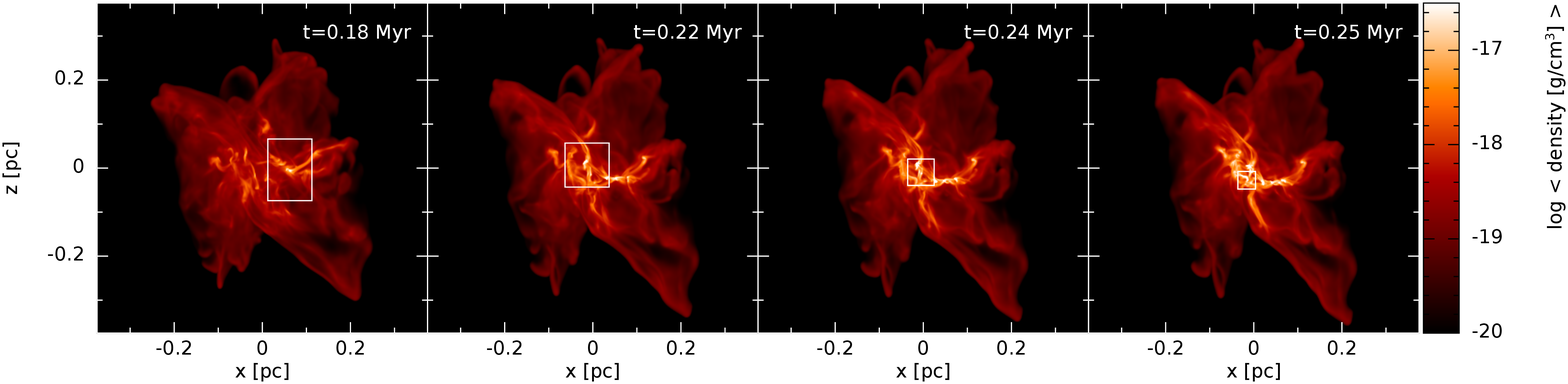}
\caption{Average gas density for four example clumps (top) and their parent cloud of \nfive{} at extraction time (bottom).  Clump masses are listed in the top row; the cloud (bottom row) has a mass of 50~\Msun{}.  The boxes in bottom row correspond to the regions that were extracted, and are the same size as the reference boxes in the top row.  We use average gas density rather than column density since there is not a consistent column depth between panels.  There is not a one-to-one visual correspondence between the two rows since the bottom row includes additional gas in the $y$-direction used to calculate the average density but not used in the extraction; i.e., projection effects.  Clearly, initial clumps are already evolved, and cannot be represented by the conventional initial conditions of uniform or centrally-condensed spheres.}
\label{fig:morph}
\end{figure*} 
Each clump has already undergone considerable evolution since the beginning of the parent simulation and is well-structured.  There is a variety of clumps in our suite, ranging in size and amount of structure.  The large clumps from the hydrodynamic cloud appear turbulent with random density fluctuations, while the clumps from strongly magnetised clouds already have filamentary structures and coherent flows.  Thus, there is diversity of clumps from both within a parent cloud and amongst the parent clouds. This reinforces that idealised initial conditions, such as uniform density or supercritical Bonner-Ebert spheres\footnote{These are the common initial density profiles as used in the studies listed in the Introduction.} are poor initial cores, even if they are turbulent.  Given the extent and structure of example clouds, it is clear that our clumps may contain cores, but themselves are not cores.  Given the extent of the clump and its filamentary structure, this suggests that gravity alone will not be responsible for star formation/evolution and suggests the inertial-inflow model  \citep{Padoan+2020} is valid for low-mass star formation as well.   

Each parent simulation contains at least two massive clumps since each simulation has one or two star-forming hubs with many stars.  Thus, the gas is associated with several stars (both young and old as discussed above), leading to a massive clump for each star.  Therefore, many of the massive clumps contain the same gas particles -- both associated and augmented gas, particularly \hyd{}.  The exceptions are \ifive{} and \nfive{} which have no massive clumps.   Rather than forming one or two massive hubs, these simulations formed many smaller hubs of only a few stars each.  The smaller hubs and lower multiplicity yield less associated gas with each star and hence less total mass.  Similarly, all star formation in \nthree{} occurred in two hubs (except the young star that formed at \sm255~kyr), hence all but the youngest star yields a clump with $M > 3$~\Msun{}.

Therefore, since star formation proceeds in (large and small) hubs, many of these clumps better represent the progenitor to systems rather than to individual stars.  This is reinforced when we look at the overlap in the gas between various clumps.  For example, at \tfinal{}, there is a tight binary in \nfive{} that is part of a quadruple system.  The two clumps originating from the binary have augmented masses of 1.81 and 1.61~\Msun{}, respectively.  There is 1.60~\Msun{} that is common to both clumps, which is 88.5 and 99.2 per cent, respectively.  This suggests that the more massive clump may reproduce the binary system.  The other two stars in this quadruple system have recently been captured, thus including them in constructing a common clump would have created an excessively large clump that is counter to our goal and possibly even unnecessary to reproducing the binary; see \citetalias{WursterRowan2023b} for further discussion.  Therefore, multiplicity is typically and implicitly accounted for in our method.  We explicitly note that this conclusion is valid only when considering the augmented clumps; for the associated clumps, the per cent overlap of our example pair decreases to 34.7 and 73.0, respectively.  

Given the highly structured clumps, we do not attempt to fit 1D profiles, given how much information will be lost.  However, this may be the focus of a future study.  

\subsection{Extraction and evolution times}
The clump extraction time is shown in the top panel \figref{fig:cloudtimes}.  
\begin{figure} 
\centering
\includegraphics[width=\columnwidth]{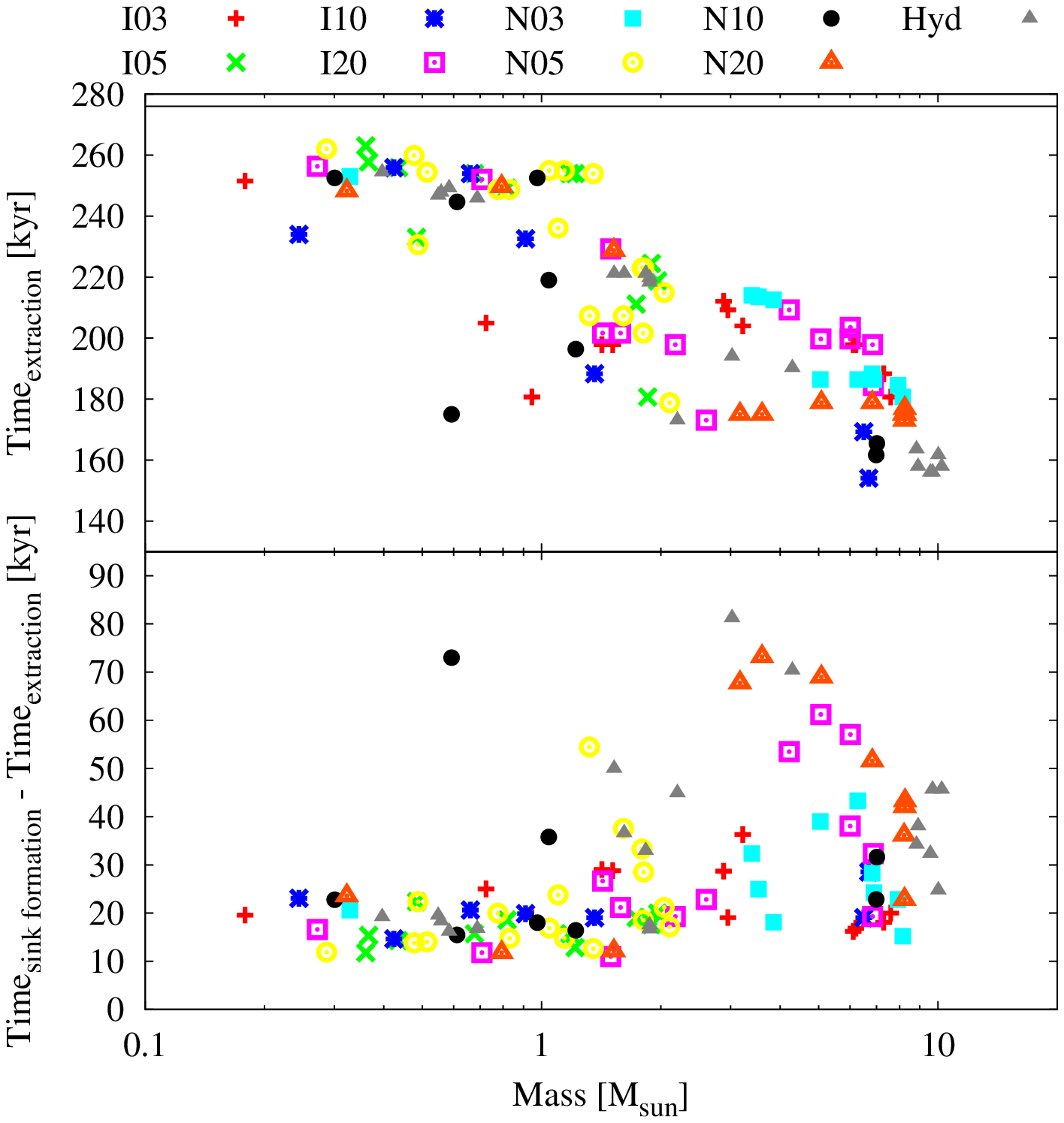}
\caption{\textit{Top}: The time in the parent simulation that each clump was extracted; the horizontal line at 276~kyr represents the end of the parent simulations.  \textit{Bottom}: The time after extraction that the sink formed in the parent simulation.  There is a general tend that more massive clumps are extracted at earlier times than lower mass clumps. The majority of the clumps evolve for \sm20~kyr before forming a sink.}
\label{fig:cloudtimes}
\end{figure} 
Clumps are extracted throughout the parent simulations, reflective of the on-going star formation.  There is a general trend that more massive clumps are extracted earlier in the parent simulation, starting at \sm155~kyr \appx{} 0.84~\tff{}.  Clumps are typically extracted \sm20~kyr before the stars forms, although in some cases, this can be upwards of \sm80~kyr, as shown in the bottom panel of \figref{fig:cloudtimes}.  In general, this is independent of the initial magnetic field strength of the parent cloud or the inclusion of non-ideal MHD.  

The early-forming massive clumps quickly form stars.  This is reasonable since these first-born stars have a long time in the parent simulation to interact with gas and other stars to build up their associated clump.  However, there are some late forming stars that also originate from massive clumps; these stars also have a longer time between clump extraction and star formation (e.g., \itwenty{} and the diffuse clumps). These stars are members of higher-order systems, thus the associated gas was already in the system for a considerable length of time due to longer-lived companions.  Although only the gas currently in the system would be associated with the younger star, it would have likely travelled from a great distance due to the companions; a greater amount of augmented gas would hence be required.  Indeed, these stars from massive clumps that form late have lower ratios of associated-to-augmented clump masses than the early forming massive clumps.  This confirms that these stars interacted primarily with the gas already in the system whereas the older companion stars interacted with additional gas throughout their lifetime prior to forming the higher-order system.

Given multiplicity at the final time, if we instead consider the difference between the extraction time and the formation of the first star in the system (as determined at the final time), we find that the difference in time typically decreases to \sm20~kyr.  However, there are some clumps that are extracted \textit{after} the first star in the system has formed.  These systems typically become bound late, and the early evolution of the gas is somewhat independent of one another.  Therefore, the trend remains that clumps are typically extracted \sm20~kyr, while notable outliers exist. 

\subsection{Clump properties}
\label{sec:clump}
Given that our parent simulation employs a Lagrangian method, all properties (expect size) are mass-weighted values.  

\subsubsection{Sizes and shapes}
Unlike the typical cores used for isolated star formation models, our augmented clumps are irregular shape (c.f., \figref{fig:morph}).  To characterise their shape, we use the ellipse fitting algorithm from \citet{WursterBonnell2023}, which is based upon \citet{RochaVelhoCarvalho2002}.  \figref{fig:cloudshape} shows the length of the semi-major axis and the ratios of the three axes.  
\begin{figure} 
\centering
\includegraphics[width=\columnwidth]{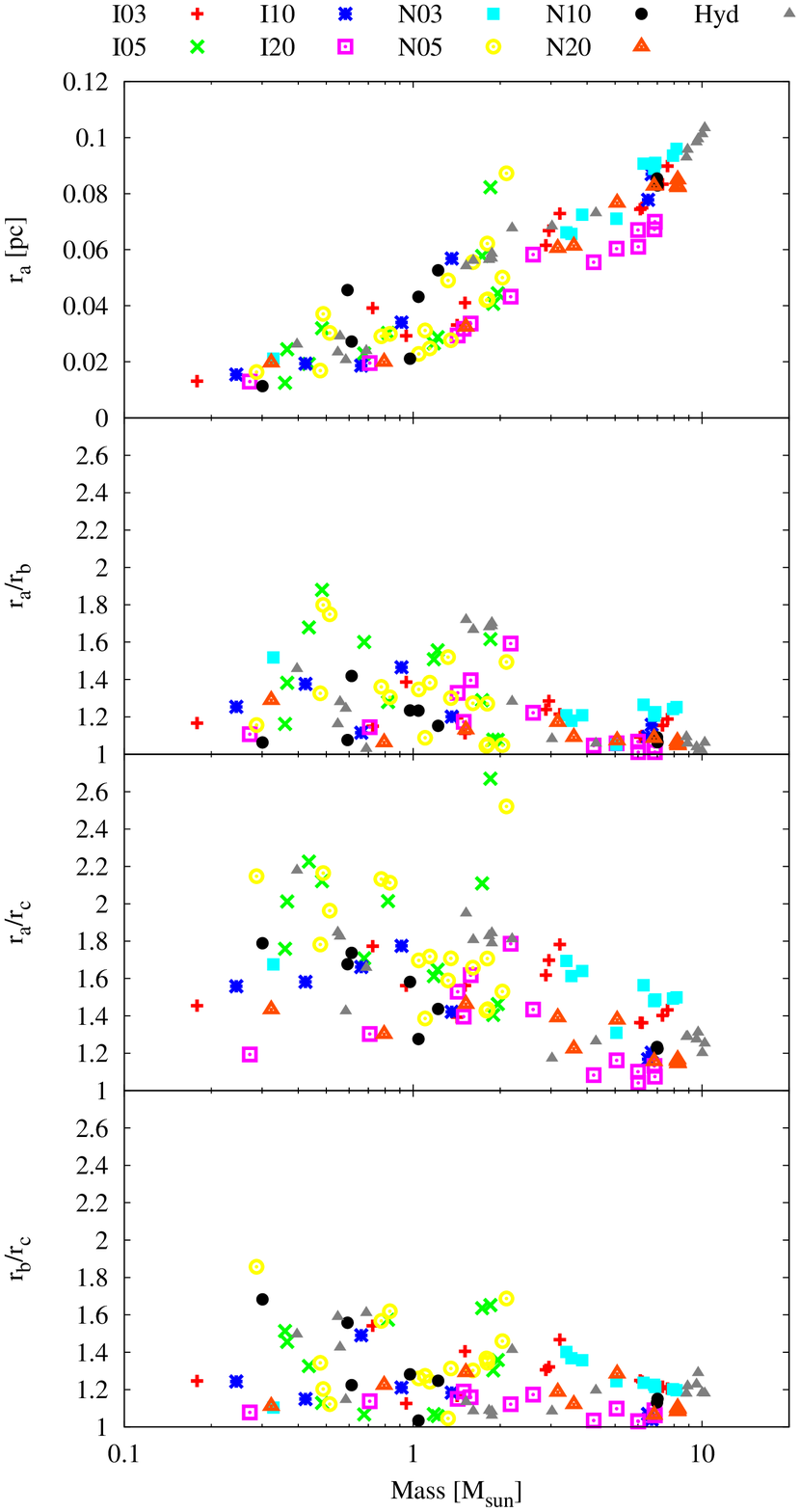}
\caption{\textit{Top}: The semi-major axis of the best-fit ellipse that characterises the clump.  \textit{Remaining panels}: ratio of the axes, where $r_\text{a} > r_\text{b} > r_\text{c}$.  Naturally, more massive clumps are larger, with maximum semi-major axis $r_\text{a}$ being slightly more than half the initial radius of the parent cloud.}
\label{fig:cloudshape}
\end{figure}

More massive clouds have larger semi-major axes (top).  The parent cloud has an initial radius of $r_0 = 0.1875$~pc, thus the most massive clumps have semi-major axes of \sm0.3-0.6$r_0$.    These massive clumps also tend to be more spherical, with the ratio of the axes $1 \lesssim r_i/r_j \lesssim 1.5$.  Most of these clouds are extracted at an early time before a well-defined filamentary network has formed (most notable in \ntwenty{} and \hyd{} as shown in \figref{fig:morph:all}), therefore, the initial gas is somewhat isotropically located within the parent cloud.  The lower mass clumps tend to be more triaxial, therefore, the gas has already evolved to form filaments along which gas can easily flow.  Given that these lower mass clumps are not strongly elongated suggests that the initial gas is not all contained in the filaments, and much gas has yet to accrete onto the filaments.  Therefore, well-defined filaments alone do not contain the entire reservoir of gas for a star.  The shape does not depend on the initial magnetisation of the parent cloud.  

Recall that to generate these clumps, we tested all particles within $r_\text{max} = 1000$~au $ = 0.0048$~pc; therefore, in all cases, the gas has flowed in from distances much greater than this.  Given the discs at the end of the parent simulations \citepalias[see fig. B1 of ][]{WursterBatePrice2019}, the bound structures appear to be $r_\text{bound} \lesssim 200$~au, therefore, $r_\text{max} = 1000$~au is a very liberal parameter.  In the literature, many simulations of isolated star formation are initialised with \sm1~\Msun{} cores of radius 0.013-0.04~pc \citepeg{Joos+2013,Tomida+2013,Tsukamoto+2015hall,WursterPriceBate2016,WursterBatePrice2018sd,Marchand+2020}, which is reasonably consistent with the range of our \sm1~\Msun{} clumps.

\subsubsection{Gas density}
As shown in \figref{fig:morph}, the extracted clumps are structured with a range of densities.  \figref{fig:clouddensityA} shows the log-averaged density\footnote{i.e., $\bar{\rho} = 10^{\left(\sum_i^N \log \rho_i\right)/N}$} of each clump and the range encompassing 95.4 per cent of the densities\footnote{The range is centred on the median, not the mean, to highlight any asymmetries about the mean.}.
\begin{figure} 
\centering
\includegraphics[width=\columnwidth]{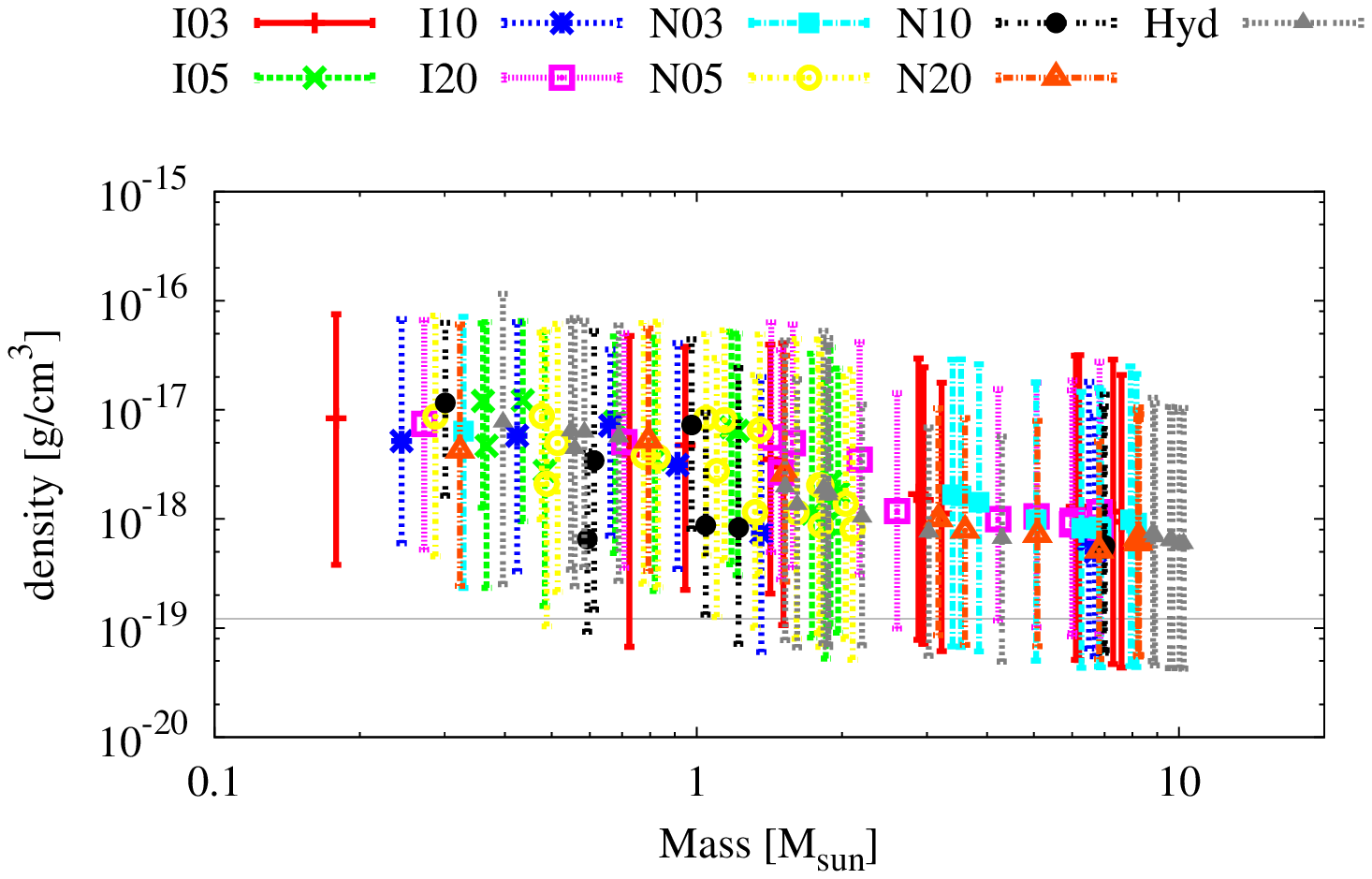}
\caption{The log-averaged gas density for each clump (symbols), where the bars bracket 95.4 per cent of the densities centred on the mean. The horizontal line represents the initial uniform density of the parent cloud.  There is a weak trend of decreasing average density with increasing clump mass.}
\label{fig:clouddensityA}
\end{figure} 
The range of densities within a clump is $\gtrsim2$~dex, which is larger than the typical spread of a factor of \sm20 in a supercritical Bonnor-Ebert sphere.  The range is bounded on the upper end by $\rho_\text{ext} = 10^{-16}$~\gpercc{}; although the initial density of the parent cloud is \rhotwoeq{1.22}{-19}, the gas has evolved such the some of the extracted gas has decreased below this value, leading to no hard lower limit.    

The clump is extracted when the densest particle has a density of \rhols{-16}, however, the log-average gas density is \sm1.5-2~dex lower.    As clump mass increases, there is a general shift to lower average gas densities.  This is reasonable given that there is only a small region in each clump of high density gas near the extraction threshold (recall  \figref{fig:morph}) and the rest is more disperse; for less massive clumps, this denser region would comprise a relatively larger fraction of the total cloud, increasing the log-average value. 

\figref{fig:clouddensityA} suggests that the mean and median gas density are similar in most clumps (compare the symbol to the midpoint of the range); however, several clumps, such as the low-mass clumps in \ithree{}, have strongly skewed distributions with more high-density gas than low-density gas.  These are reasonably centrally-condensed clumps with very little diffuse gas surrounding the dense region.  These skewed distributions are highlighted in \figref{fig:clouddensityD}, which shows the normalised gas density distribution for each clump from each parent simulation.  
\begin{figure} 
\centering
\includegraphics[width=\columnwidth]{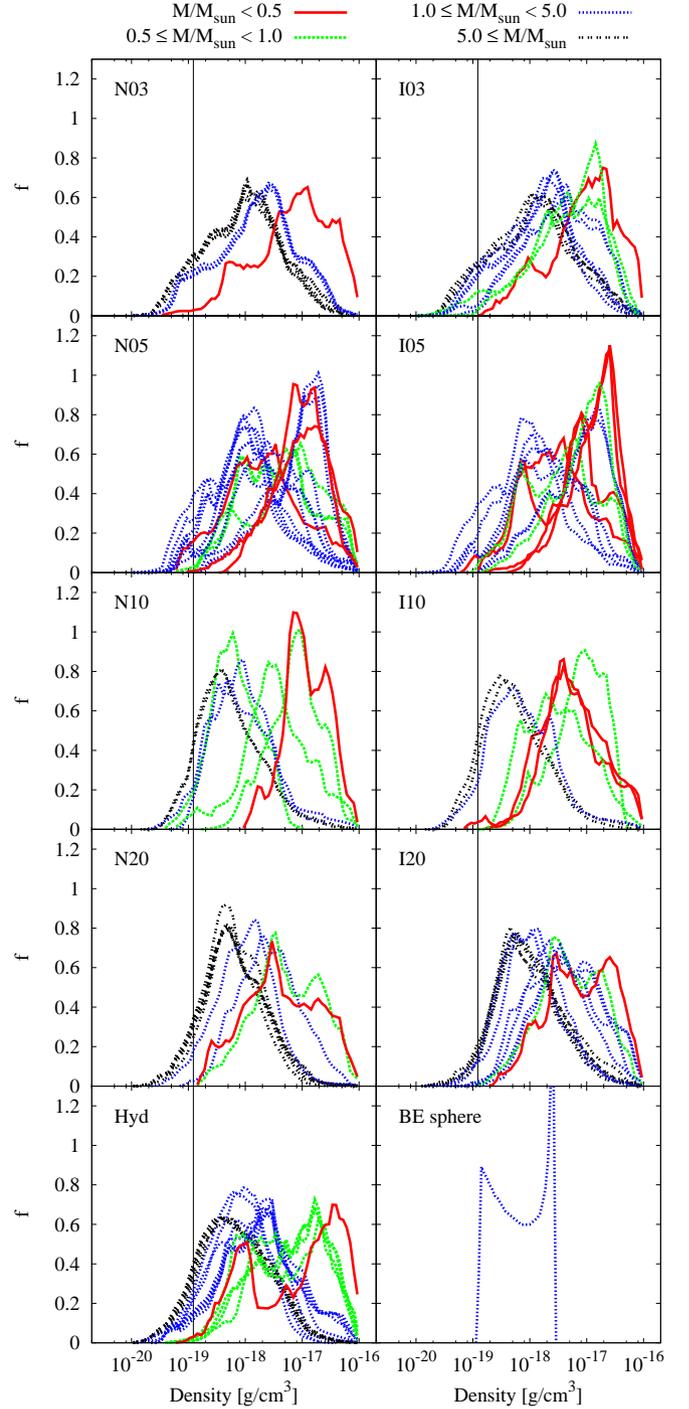}
\caption{The normalised gas density distribution of each clump.  The colours represent mass ranges of the clumps.  The vertical lines represent the initial density of the parent cloud.  There is no consistent gas density distribution, demonstrating the varied nature of star forming regions.  The bottom right panel is the normalised gas distribution of a supercritical Bonnor-Ebert sphere.}
\label{fig:clouddensityD}
\end{figure}  
There is no consistent gas density distribution.  All distributions are generally uni-modal, however, the peak of the distribution can be near the median density or skewed to higher or lower densities.  In general, higher-mass clumps have distributions skewed to lower densities while the lower-mass clumps have distributions skewed to higher densities.  Again, this is reflective of relative quantity of surrounding disperse gas.  The deviations from a smooth curve reflect the substructures that have formed away from the region that will likely collapse to for the star.  This demonstrates the chaotic and varied nature of star forming regions.  

For reference, the bottom right panel of \figref{fig:clouddensityD} shows the density distribution of a 1~\Msun{} supercritical Bonnor-Ebert sphere with a concentration parameter of $\xi = 7.45$.  This centrally-condensed sphere yields a smooth density distribution that is bi-modal, with local maxima at both extremes of density within the sphere.  This distribution is opposite that of the extracted clumps, where the single maximum is located away from the extremes.  This reinforces that we have extracted clumps and not cores.  Furthermore, the gas distributions of our clumps do not include any structures that reflect the density distribution of a Bonnor-Ebert sphere; this suggest that either a core has yet to form in our clump, or that a Bonnor-Ebert sphere is a poor representation of a star forming clumps.  To distinguish between the two options, we would need to extract the clumps at several times closer to the epoch of star formation, but that is beyond the scope of this work.   Therefore, although Bonnor-Ebert spheres are a more reasonable initial condition then a sphere of uniform density, they poorly represent a realistic star forming clump.  

\subsubsection{Temperature}
\label{sec:results:temp}
The average gas temperature does not vary much amongst the clumps, with $6 \lesssim \left<T\right>/\text{K} \lesssim 9$, as shown in \figref{fig:cloudtemp}.  This is slightly cooler than the initial parent cloud, which was initialised with $8.8 \lesssim T/\text{K} \lesssim 13$, indicating that majority of the gas has already begun to cool and condense by extraction time; this is is confirmed by the filamentary structures as in \figref{fig:morph} even at this early time.    Some of the low-density gas outside of the filaments, however, has been heated, leading to the highly skewed distribution.  However, this `hot' gas is still only a small fraction of total gas mass of the clumps, and its absolute temperature is still cool.  Therefore, the initial star forming clump is a cold object that can be cautiously represented as isothermal.
\begin{figure} 
\centering
\includegraphics[width=\columnwidth]{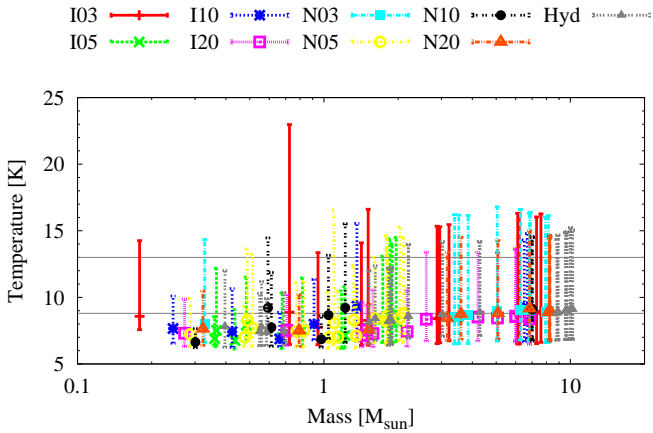}
\caption{The average gas temperature for each clump (symbols), where the bars bracket 95.4 per cent of the temperatures centred on the mean. The horizontal lines represent the bounds of the initial temperature of the parent cloud.  The gas is still cold, with the distribution skewed to lower temperatures.}
\label{fig:cloudtemp}
\end{figure} 

\subsubsection{Turbulence}
\label{sec:results:turb}
Star forming regions are observed to be turbulent \citepeg{Larson1981,DubinskiNarayanPhillips1995,HeyerBrunt2004,MaclowKlessen2004}.  Historically, Larson's relationship between mass and velocity dispersion has yielded at tight empirical correlation, and hence an excellent diagnostic for numerical simulations.  However, recent observations have called into question the  validity of Larson's Laws, suggesting that observational biases may have contributed to the relationship; see discussion and references in \citet{Traficante+2018lar}.  Nonetheless, as a diagnostic, we plot the three dimensional velocity dispersion in the top panel of \figref{fig:cloudturb}, including the empirical line of best fit from \citet{Larson1981}.
\begin{figure} 
\centering
\includegraphics[width=\columnwidth]{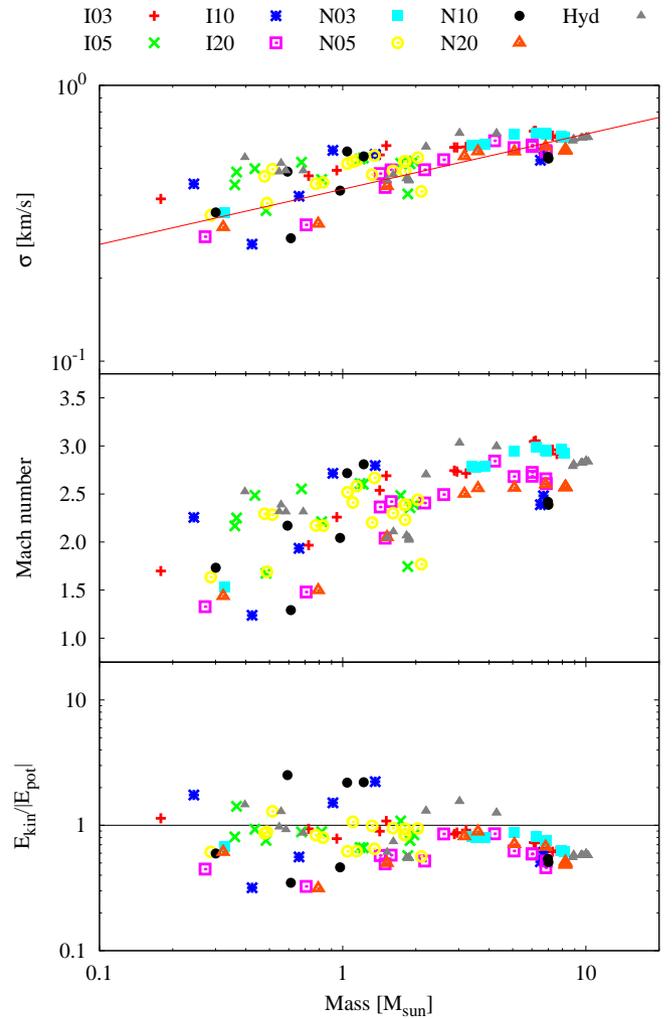}
\caption{Our star forming clumps are turbulent.  
\textit{Top}: The three dimensional velocity dispersion for each clump.  The line represent the empirical line of best fit from \citet{Larson1981}, such that $\sigma \ [$\kms{}$] = 0.42 M^{0.20}$~\Msun{}.  Our clumps have reasonable agreement with the observed relationship.
\textit{Middle}: The rms Mach number of each clump.  The initial Mach number of the parent cloud is 4.8.  Every clump is supersonic, with more massive clumps having higher Mach numbers.
\textit{Botton}: The ratios of kinetic to gravitational potential.  The ratio in the parent cloud is 1.  The majority of the clumps has a ratio less than one, suggesting that it may already be undergoing gravitational collapse.}
\label{fig:cloudturb}
\end{figure} 
We find reasonable agreement to Larson's line.  This might not be unreasonable given that all our clumps represent the same early evolutionary phase and were extracted from very similar parent simulations.  

The rms Mach number is another indication of the level of turbulence in clumps; see middle panel of \figref{fig:cloudturb}.  Every clump is supersonic, with more massive clumps having higher Mach numbers.  The Mach number of the clump does not necessarily reflect the global Mach number of the parent cloud at extraction time, as shown in \figref{fig:cloudmach}.  The initial Mach number of the parent clouds is 4.8, which decays over the first 100-200~kyr before increasing again, primarily due to the increased motion of the gas as cores form, interact, and evolve.  Some of the more massive clumps extracted at earlier times have Mach numbers higher than the parent cloud at extraction time; this is a result of the gas at the edge of the parent cloud having low a Mach number (causing a decrease in the rms value) but not being included in these massive clumps.  Most of the low-mass clumps extracted later have Mach numbers well below that of the parent cloud, showing how the turbulence has decayed as it cascades to smaller scales.  Therefore, the rms Mach number of the clumps reflects the scale of the turbulence and the level to which it decayed.

While this general trend (lower-mass cores have lower rms Mach numbers) is similar to that in \citet{Pelkonen+2021}, their clumps are generally smaller and have lower Mach numbers for clumps of similar mass.  Their initial rms Mach number is higher than ours, however, their simulation has evolved much longer before their cores were extracted, permitting additional time for their turbulence to decay.  Therefore, a clump's Mach number is likely dependent on the length of time the simulations are permitted to evolve before clump extraction.
\begin{figure} 
\centering
\includegraphics[width=\columnwidth]{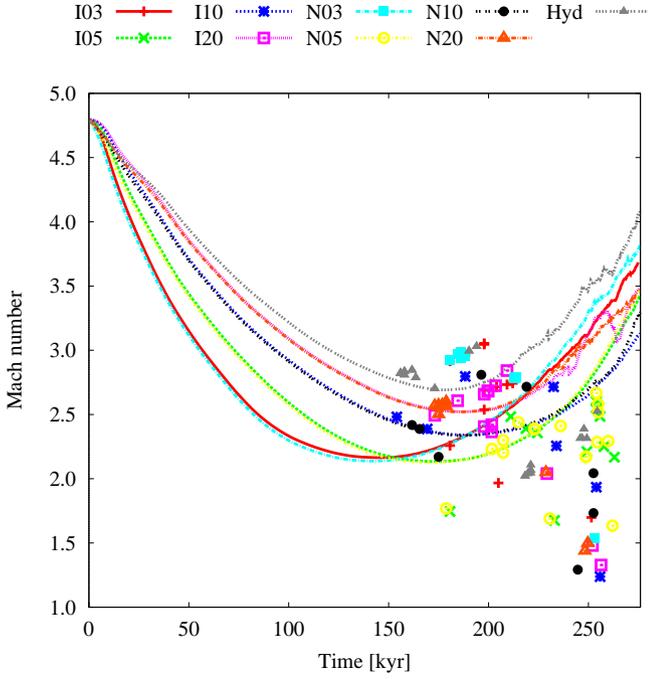}
\caption{The evolution of the rms Mach number of the parent clouds (lines) along with the rms Mach number of the clumps at extraction time (symbols).  The turbulence in the low-mass clumps has decayed such their Mach number is well below that of the parent cloud.}
\label{fig:cloudmach}
\end{figure} 

The initial Mach number of the parent cloud was chosen such that the potential and kinetic energy balance one another (i.e., $E_\text{kin}/\left|E_\text{pot}\right|  = 1$).  At extraction, the clumps have ratios of $E_\text{kin}/\left|E_\text{pot}\right| < 3$ with the majority of the clumps having $E_\text{kin}/\left|E_\text{pot}\right| \lesssim 1$, as shown in the bottom panel of \figref{fig:cloudturb}.  This is approximately independent of mass, although the massive clumps have ratios less than one.  Therefore, at extraction time, the majority of our clumps are already undergoing gravitational collapse; this conclusion is based upon a global value of the clump, although it is possible that only the dense regions themselves are undergoing collapse.  The clumps with $M \gtrsim2$~\Msun{} and $E_\text{kin}/\left|E_\text{pot}\right| > 1$ are the diffuse clumps that may require additional time for the turbulence to decay to the level where star formation can occur.  However, these plots suggest that stars can form from even low-mass, turbulent clumps, contradicting the results of \citet{LewisBate2018}.

Star forming clumps have been observed to have a bulk rotation \citepeg{Goodman+1993,Caselli+2002}.  The observed ranges of the ratio of rotational to gravitational energy, $\beta_\text{r} = E_\text{rot}/\left|E_\text{pot}\right|$, varies between studies, with the extremes yielding a range of $10^{-4} < \beta_\text{r} < 1.4$;  \citet{Goodman+1993} found typical values of $\beta_\text{r} \sim 0.02$.  Calculating the rotational energy in our clumps is a challenge since there is no obvious centre of rotation.  Therefore, the rest of this paragraph must be taken with caution.  When we calculate the rotational energy about the densest region\footnote{We use the density-weighted centre of mass of all the particles within 2$h$ of the densest gas particle.\label{footnote:com}}, we find a typical range of $0.13 < \beta_\text{r} < 1.3$, with typical values of $\beta_\text{r} \sim 0.4$.  These values lie near the upper end of the observed range, but might be artificially high since our rotational energy is degenerate with turbulent energy.

\subsubsection{Magnetic field strength}
In simulations of isolated star formation, uniform magnetic fields of constant strength are typically employed (see the studies listed in the Introduction).  This permits a simple selection of the initial mass-to-flux ratio and enforcement of $\bm{\nabla} \cdot \bm{B}_0 = 0$.  Although there are options to initialise clumps with an hourglass magnetic field \citepeg{EwertowskiBasu2013,Bino+2022}, these are equally artificial if the density profile is not consistent with one that would have caused the hourglass structure.  Neither case is indicative of our clumps.  

 \figref{fig:morphB} shows the average gas density and average magnetic field strength for four example clumps from \nfive{} overlaid with magnetic field unit vectors.  
\begin{figure*} 
\centering
\includegraphics[width=\textwidth]{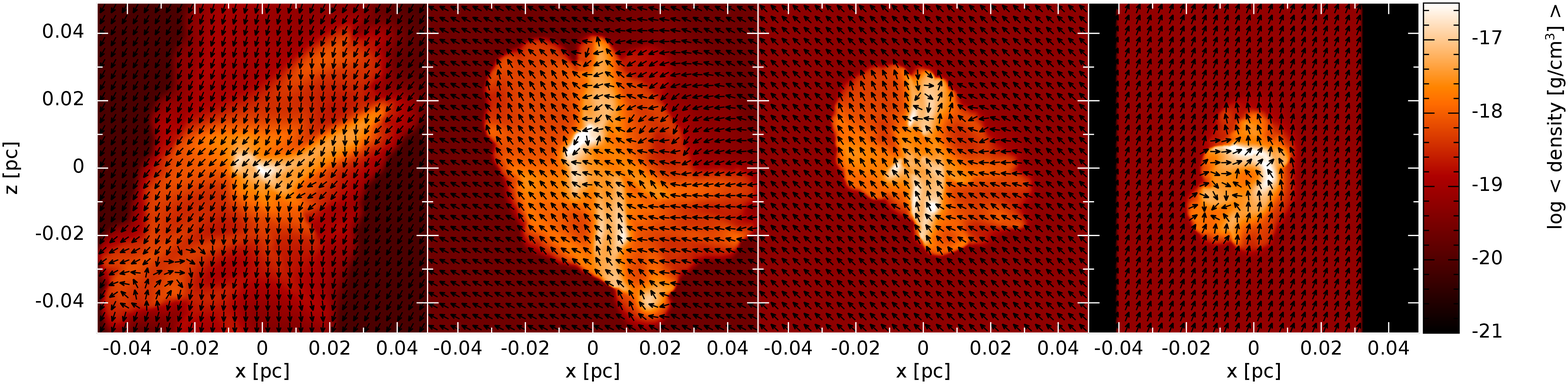}
\includegraphics[width=\textwidth]{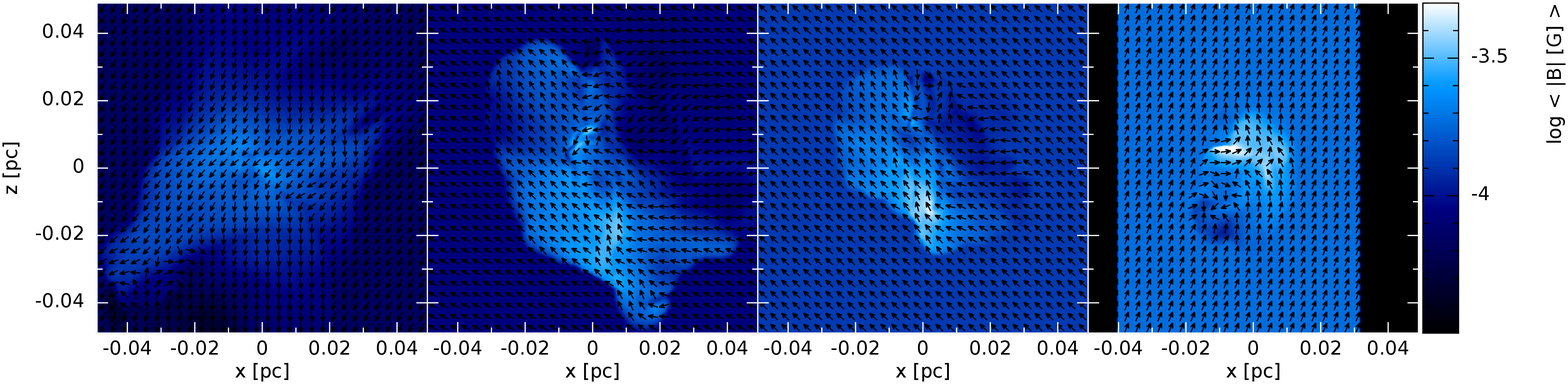}
\caption{ The average gas density (top) and average magnetic field strength (bottom) for four example clumps extracted from \textit{N05} as in \figref{fig:morph}, but with a different density scale.  The black vectors represent the magnetic field direction; the strength can be determined from the bottom plot.  At extraction time, the magnetic field strength and geometry varies throughout the clump.}
\label{fig:morphB}
\end{figure*} 
At extraction time, there is already a distribution of strengths and geometries, where the magnetic field tends to be perpendicular to the dense filaments and parallel to the less dense filaments, in agreement with observations \citepeg{Goldsmith+2008,Soler+2013,SolerBraccoPon2018,Planck2016or,Planck2016role}.

For a quantitative analysis, \figref{fig:cloudBA} shows the log-averaged magnetic field strength of each clump and the range encompassing 95.4 per cent of the field strengths.
\begin{figure} 
\centering
\includegraphics[width=\columnwidth]{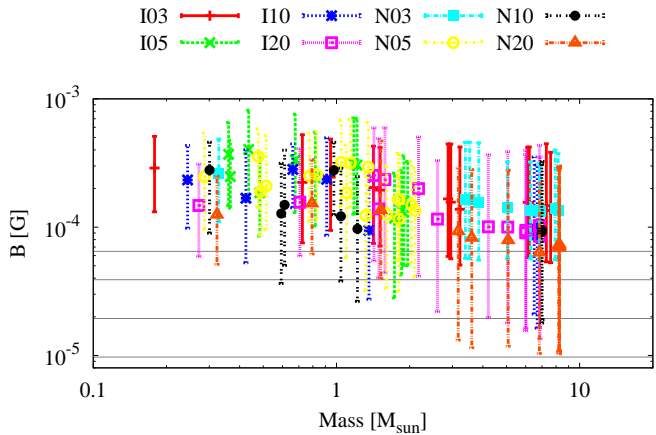}
\caption{The log-averaged magnetic field strength for each clump (symbols), where the bars bracket 95.4 per cent of the field strengths centred on the mean.  The horizontal lines represent the field strengths of the initial parent cloud.  All clumps have stronger average field strengths than their parent cloud.  There is a smaller range between the average clump strengths than between the parent cloud strengths, suggesting a reduced dependence on the properties of the parent cloud.}
\label{fig:cloudBA}
\end{figure} 
In all clumps, the average magnetic field strength has increased from the initial strength of the parent cloud, with greater increases in clumps from clouds with initially weaker field strengths.   There is a slight trend to lower averages for more massive clumps, but this is reasonable since we expect the augmented gas (in general) to have lower densities than the associated gas and hence lower field strengths.  The initial strengths in the literature typically span a larger range than in our clumps, but those strengths are based upon low mass-to-flux ratios, thus dependent on the size and mass of the idealised clump.  

When collectively analysing all the clumps from each parent cloud, there is a weak trend that parents clouds of lower initial field strengths yield clumps with lower average field strengths; the clumps in \ifive{} and \nfive{} break this trend.  This decrease of average clump strength from \itwenty{} to \ithree{} (or \ntwenty{} to \nthree{}) spans less than the factor of \sm7 that spanned the initial parent clouds.  This suggests that the clumps have only a weak dependence on the field strength of the larger environment (in this case, the parent cloud).  When considering the range of averages for clumps from each parent cloud, the maximum average from all clouds is similar, whereas the minimum average is higher for clouds with initially stronger field strengths.  The latter suggests that the parent clouds impose a floor on the magnetic field strength.  

Independent of parent cloud, the clump magnetic field strengths span a common range of $6\times10^{-5} \lesssim \left<B_\text{clump}\right>/\text{G} \lesssim 4\times10^{-4}$ \citep[in reasonable agreement with, e.g.,][]{Crutcher+2010,Eswaraiah+2021,Kwon+2022}.  This analysis of the average magnetic field strength suggests that, in star forming clumps, there is only a weak dependence on the initial field strength of the parent cloud.  

\figref{fig:cloudBD} shows the normalised distribution of magnetic field strengths for each clump.  
\begin{figure}
\centering
\includegraphics[width=\columnwidth]{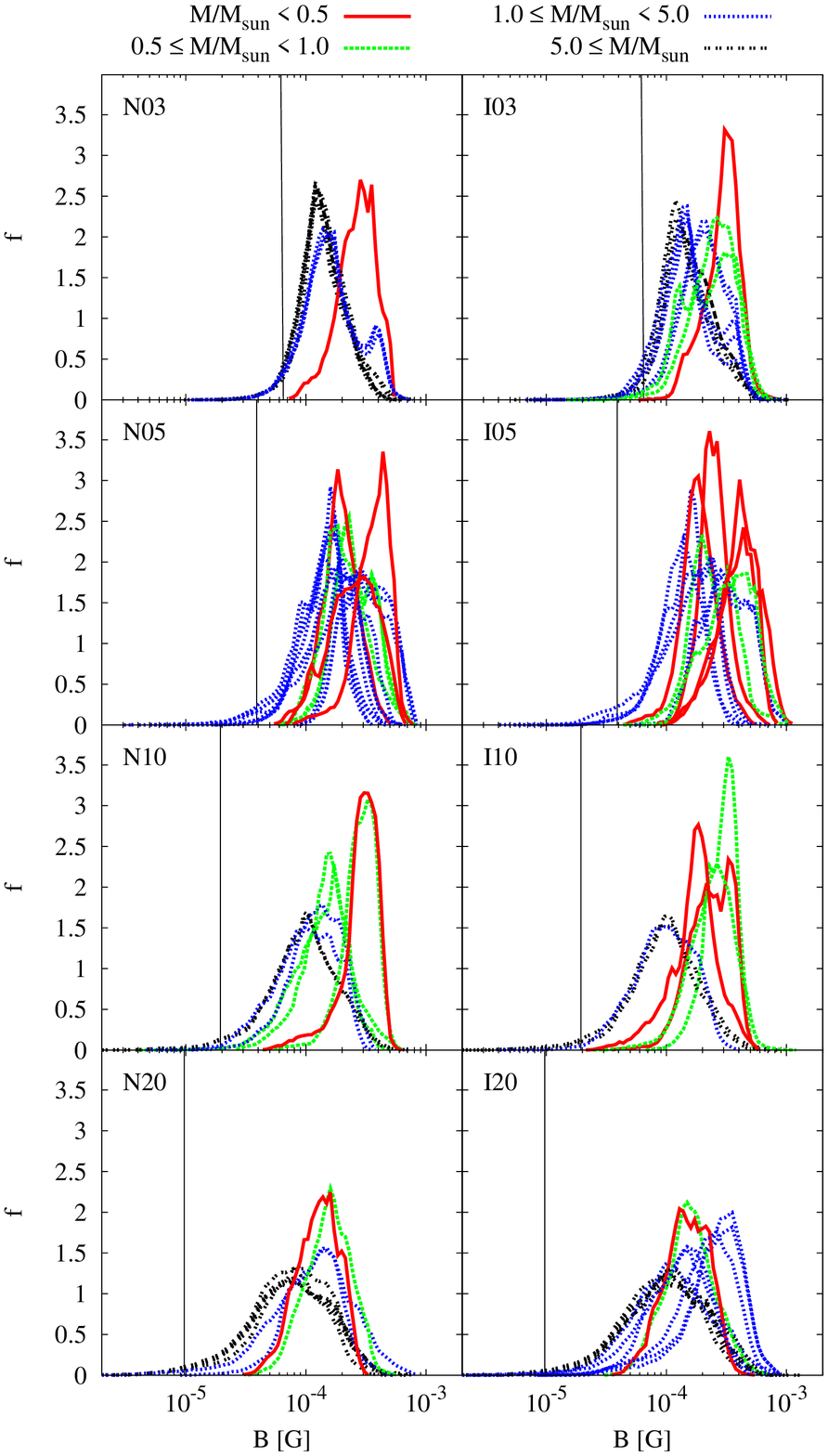}
\caption{The normalised magnetic field strength distribution for each clump.  The colours represent mass ranges of the clumps.  The vertical lines represent the initial magnetic field strength of the parent cloud.  The distributions tend to be skewed to higher strengths for lower mass clumps.  There is no clear trend in the distributions amongst the varying initial field strengths (i.e., comparing the rows), although there is consistently very little gas in the clumps with field strengths lower than the initial strength.  }
\label{fig:cloudBD}
\end{figure}
Like the gas distributions (\figref{fig:clouddensityD}), there is no consistent magnetic field strength distribution.  All distributions are generally uni-modal and span at least an order of magnitude in strength.  The peak of the distribution can be near the median strength or skewed to higher strengths; there is greater skewing for lower mass clumps.  Nearly all distributions have some gas with strengths less than the initial field strength, but this is reasonable given that some of the gas in the clump has a lower density than the parent cloud's initial density and there it a general correlation between density and magnetic field strength \citepeg{BateTriccoPrice2014,Tsukamoto+2015oa,WursterBatePrice2018sd}.  Unlike the gas distribution, these curves tend to be smoother, suggesting there is still some ordering of the magnetic field (see also \secref{sec:Btheta}).  

It is often useful to characterise a clump in terms of the normalised mass-to-flux ratio for an understanding of the relative importance of the magnetic field and gravity.  However, the ratio is geometry-dependent \citepeg{MouschoviasSpitzer1976,NakanoNakamura1978,Mestel1999,Tritsis+2015}, and is non-trivial to calculate in our structured clumps; furthermore, a global value likely does not represent the relative importance in the first region to collapse since the clump includes all the gas that ultimate interacts with the star over its lifetime.  Instead, \figref{fig:cloudbetaB} shows the ratio of magnetic-to-gravitational potential energy; this is still a global quantity so might not well-represent the first region to collapse, but it is not geometry-dependent.
\begin{figure} 
\centering
\includegraphics[width=\columnwidth]{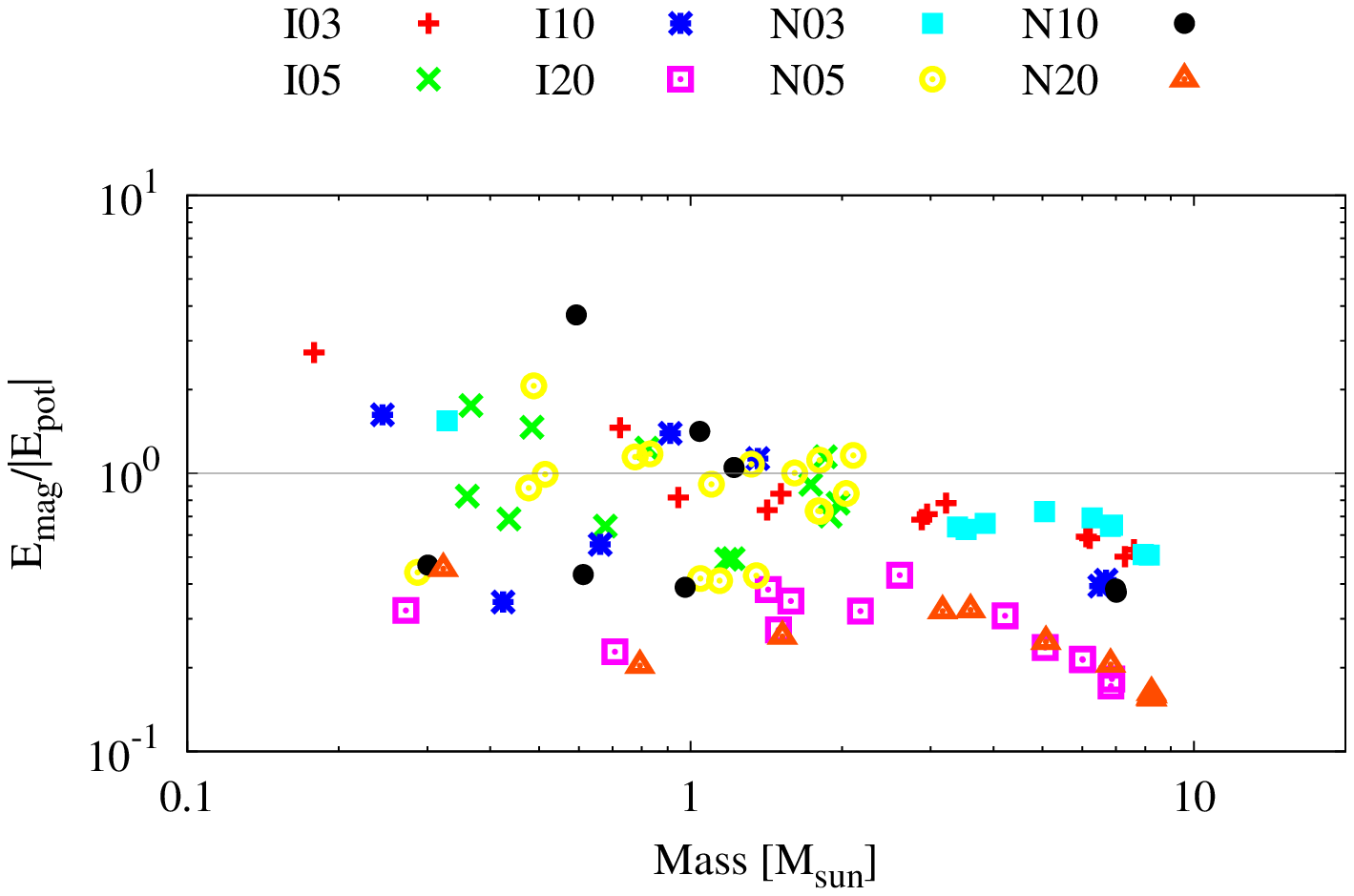}
\caption{The ratio of magnetic to gravitational potential energy.  Most clumps are gravitationally dominated, including all clumps with $M \gtrsim 2$~\Msun{}, indicating that there is not enough magnetic support to prevent star formation.  Only \itwenty{} has no magnetically dominated clumps.}
\label{fig:cloudbetaB}
\end{figure} 
Our suite includes both gravitationally dominated and magnetically dominated clumps, however, the majority of the clumps are gravitationally dominated, indicating that they are undergoing or about to undergo collapse.  Lower mass clumps are either gravitationally or magnetically dominated, while higher mass clumps tend to be gravitationally dominated, with no clump with $M \gtrsim 2$~\Msun{} being magnetically dominated.

There is no clear trend between the ratio of magnetic to gravitational energies and time, with \sm70 per cent of the clumps (both gravitationally and magnetically dominated) forming their sinks within 25~kyr of the extraction time (where the sink formation time was also extracted from the parent simulation).  This suggests that the magnetically dominated clouds either have a mechanism to quickly diffuse the magnetic field to permit gravitational collapse, or the global value is not representative of the small region that will first undergo collapse.

The ratio has little dependence on the inclusion of non-ideal MHD, indicating that, at the times and scales that the clumps are extracted, the non-ideal processes have played a minimal role.  This agrees with the conclusion in \citetalias{WursterBatePrice2019} that non-ideal effects are primarily important on small scales.

\subsubsection{Magnetic field orientation}
\label{sec:Btheta}
As shown in \figref{fig:morphB}, there is a range of magnetic field orientations in each clump.  In our examples, it is clear that the field has little memory of the initial orientation, which was anti-aligned with the $z$-axis.  This is most obvious in the background media which is reflective of the boundaries of the clumps.  To determine how much the field has evolved from its initial anti-alignment, \figref{fig:cloudThetaA} shows the average cosine of the angle between the magnetic field vector and the $z$-axis\footnote{We analyse the value of $\cos{\theta}$ since analysing $\theta$ itself yields a bias towards the field being perpendicular to the $z$-axis.}.  In the parent simulations, $\cos{\theta} = -1$.
\begin{figure} 
\centering
\includegraphics[width=\columnwidth]{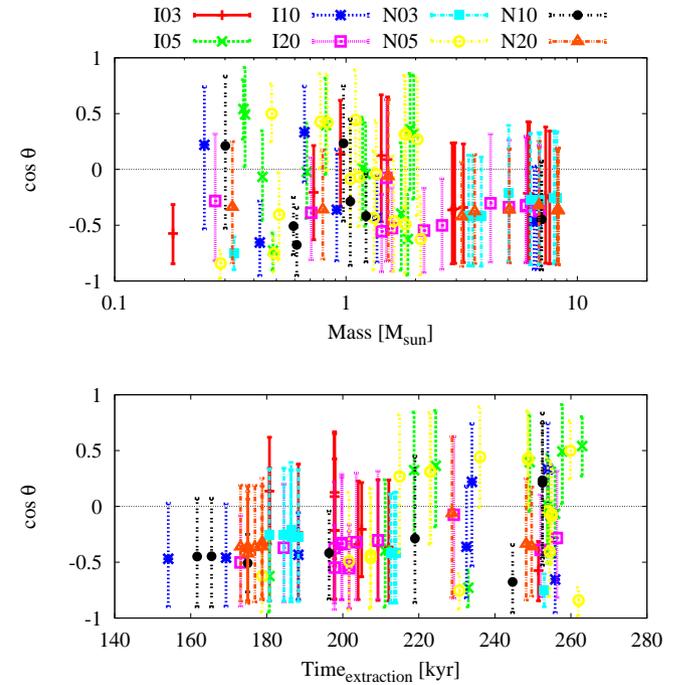}
\caption{The average angle between the magnetic field and the $z$-axis (symbols), where the bars bracket 68.2 per cent of the angles centred on the mean; the bars cover a lower spread than the previous figures, otherwise they would extend throughout most of the range.  The parent clouds were seeded with $\cos\theta = -1$.  The clumps contain only a weak memory of the field from the parent simulation, although in some late-forming clumps, this memory has been completely removed (i.e., those late-forming clumps with $\left<\cos\theta\right> > 0$.) }
\label{fig:cloudThetaA}
\end{figure} 
The spread of angles in each clump and an average away from $\cos\theta \approx -1$ suggests that the magnetic field is already twisted, as already demonstrated in \figref{fig:morphB}.  This suggests that turbulent or partially turbulent initial magnetic fields as in \citet{GerrardFederrathKuruwita2019} might yield more realistic initial conditions.  

In several low-mass clumps, the magnetic field has flipped orientation with respect to the $z$-axis, yielding $\left<\cos\theta\right> > 0$ (see also the right-most panel in \figref{fig:morphB}); these clumps form later in the parent simulations, after the magnetic field has already been twisted and affected by the older stars (bottom panel of \figref{fig:cloudThetaA}).  Therefore, most clumps retain only a slight memory of the primordial magnetic field, although the initial field geometry has been completely washed out for some late-forming clumps. 

For a better analysis of the spread of angles, \figref{fig:cloudThetaD} shows the normalised distribution of the angles.  
\begin{figure} 
\centering
\includegraphics[width=\columnwidth]{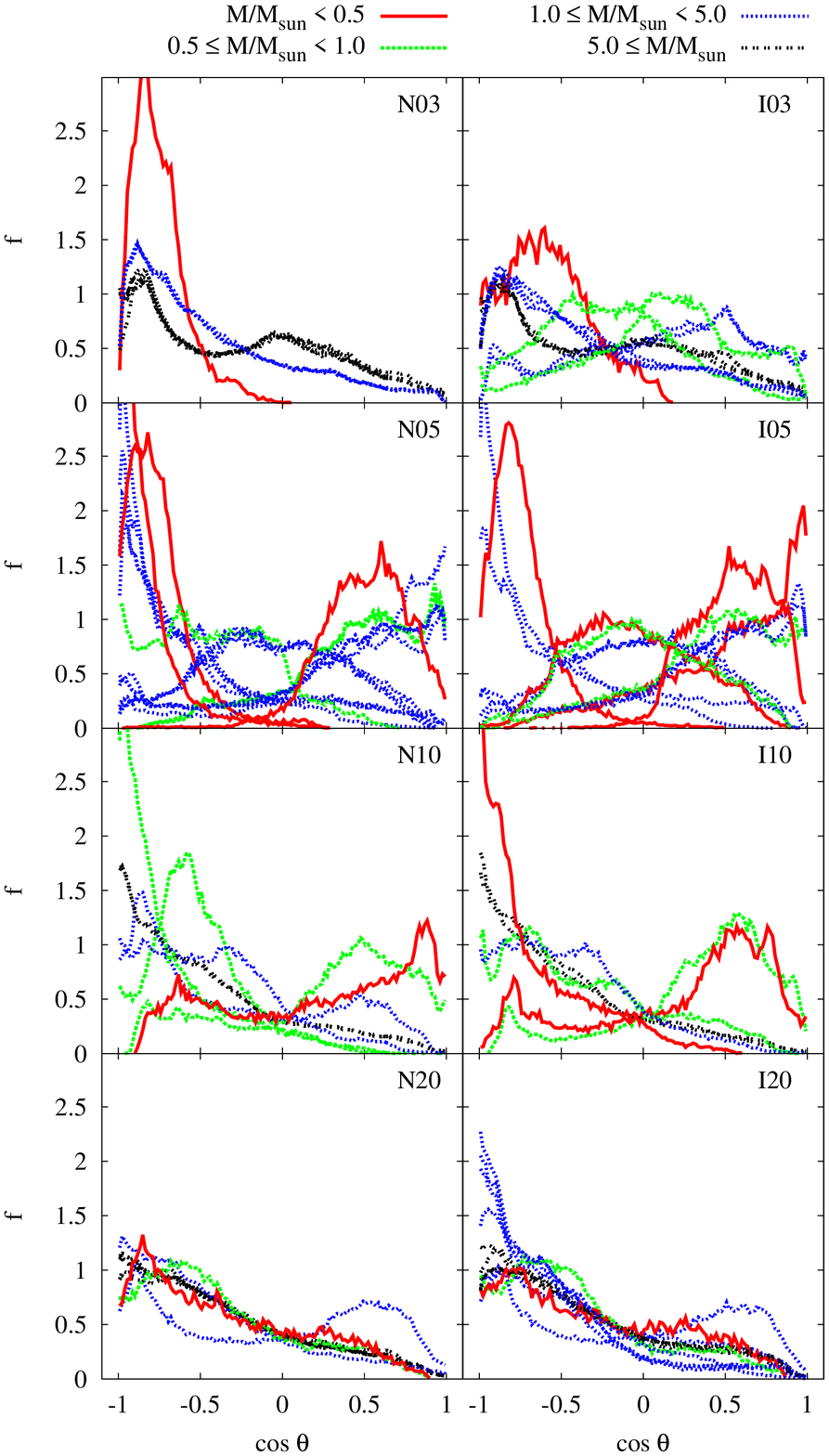}
\caption{The normalised distribution of angle between the magnetic field and the $z$-axis for each clump; the angle of the parent clouds was $\cos\theta = -1$.  The colours represent mass ranges of the clumps.  A few models have $f > 3$ near $\cos\theta \approx -1$, but we truncate the vertical axis for clarity of the distribution at smaller $f$.  Most distributions cover the entire permitted range of angles, indicating an evolved and twisted magnetic field structure.}
\label{fig:cloudThetaD}
\end{figure} 
For nearly every clump, the distribution of $\cos\theta$ covers the entire range of possible values, although, in most clumps, there is a preference for lower values.   Unlike the density or magnetic field strength, most of these distributions are not strongly peaked; moreover, some clumps have weakly bimodal distributions, which might indicate the presence of high- and low-density filaments \citepeg{Goldsmith+2008,Soler+2013,SolerBraccoPon2018,Planck2016or,Planck2016role}.  There are a few clumps in \nthree{}, \ifive{}, \nfive{}, \iten{}, and \nten{} that have strongly peaked distributions near $\cos\theta \sim -1$; these are not early-forming clumps, thus we cannot be sure if these clumps have actually retained a memory of the initial magnetic field, or if the field has evolved such that its configuration again matches the primordial field.  

Therefore, the magnetic field in a star forming clump is twisted with a complex geometry.  There is no consistency of orientation or geometry between clumps, even those that are spatially and temporally correlated from the same parent simulation.  This is consistent with observations of Taurus, where three nearby clumps have different magnetic field strengths and orientations \citep{Eswaraiah+2021}.

A more useful analysis would be to compare the magnetic field vector with the angular momentum vector, since misalignment hinders the transport of angular momentum.  However, as discussed in \secref{sec:results:turb}, determining the point about which there is coherent rotation is challenging, if such a point even exists.  Assuming the origin is near the densest particle (see footnote \ref{footnote:com}), the average angle is $\left<\cos\theta\right> \approx 0$ with 95.4 per cent of the gas spanning nearly the entire permitted range of  $\cos\theta \in \left\{-1,1\right\}$.  This represents a random distribution, suggesting either no preferred alignment between the magnetic field and the local angular momentum vectors, or, more likely, that the clump has yet to form a coherent rotation.

\section{Discussion}
\label{sec:disc}
\subsection{Caveats}

Although we present star forming clumps in this paper, their distribution of masses and radii are consistent with the star forming cores used in the literature to model the formation of an isolated star or a binary system.  Therefore, our clumps are better poised as more realistic initial conditions for such simulations.   However, we acknowledge a few caveats regarding our above results.  

All nine parent clouds were only 50~\Msun{} and had the same initial conditions, except for the magnetic field strength and the inclusion/exclusion of non-ideal MHD.  The relatively low mass and high density of the cloud promoted the formation of low-mass stars (at $t_\text{final}$, the most massive star was 2.23~\Msun{}) and the dense environment promoted the formation of multiple systems.  Therefore, our clumps will have a bias to reproduce a similar stellar mass range and multiplicity.

Next, we are unsure how much the clumps retain of a memory of the cloud's initial conditions.  Although ratios of $E_\text{kin}/\left|E_\text{pot}\right| \lesssim 1$ seem reasonable, is our range of Mach numbers (c.f. middle panel of \figref{fig:cloudturb})?  Our clumps are all supersonic, while \citet{Pelkonen+2021} obtained both super- and sub-sonic cores.  Although we found a correlation between Mach number and clump mass, will this relationship change if the parent cloud was seeded with a higher or lower Mach number or if the cloud evolved longer before extraction time?  Given the minimal impact of changing the magnetic field strength and processes, it is possible that changes to the initial Mach number may also have a minimal impact on the clump, but this is speculation.  

Ideally, to determine the impact of the properties of the initial parent cloud, we would perform additional cluster simulations varying the initial mass, density, mach number, etc.... However, each simulation in \citetalias{WursterBatePrice2019} took up to 1.5~yr of wall time\footnote{Closer to 2~yr when accounting for time spen in the queue and maintenance time of the cluster.} and \sm$10^6$ CPU hours, thus is it not practical to perform additional parent simulations at this time, particularly with greater masses.  In the future, as computers improve, then we may be able to perform a more comprehensive suite of parent simulations to determine how dependent the clump properties are on the parent cloud.

Within our current study, defining later  $t_\text{final}$ (when possible) or lower $\rho_\text{ext}$ yield larger clumps.  The former permits the star to interact with more gas during its longer lifetime, thus increasing the mass of the associated clump.  The latter requires the gas to be extracted at an earlier time when the associated gas is more disperse; this requires more augmented gas than compared to higher values of  $\rho_\text{ext}$.  In the current study, $t_\text{final}$ was chosen to be the maximum possible time consistent amongst our suite of parent simulations, thus we cannot choose a larger value.  As discussed in \secref{sec:ext:ext}, any extraction density with $\rho_\text{ext} \lesssim $\rhoxapprox{-15} should be valid.  We tested multiple values, and ultimately chose $\rho_\text{ext} = 10^{-16}$~\gpercc{} since this is high enough to guarantee star formation but not so high as to seed the clump with a first hydrostatic core, and not so low as to form massive clumps that would require considerable computational expense to evolve (as in our planned follow-up studies).  Additionally, the average gas density in the clump (\figref{fig:clouddensityA}) is reasonable compared to simulations of isolated star formation.  

While we acknowledge that there are caveats and free parameters associated with this study, the clumps presented here are nonetheless more realistic than those cores commonly used in the literature for studies of isolated star formation.

\subsection{Diffuse clumps}
\label{sec:diffuse}
In \secref{sec:ext:aug}, we enforced that 90 per cent of the associated particles must be retained in the final clump.  This was to ensure that the majority of the gas that would interact with a star existed in its clump.  In most clumps, this threshold was automatically reached by our neighbour search.  However, for progenitor sinks in some higher-order systems, our neighbour search found only a small fraction of the associated particles.  In these clumps, much of the gas was likely associated with another star in the system for most of its life, but then became associated with the progenitor late in the simulation; therefore, although associated with the star, much of this gas likely played no role in the star's actual formation or early evolution.  Since this gas likely came in from far away, it requires a considerable amount of augmented gas (assuming the 90 per cent threshold is retained), leading to the low fraction of associated-to-augmented gas mass.

Although this associated gas technically meets our criteria, should it be included in the augmented clump if is not identified at extraction time through our neighbour search?  If we include this gas, then the clump will represent a near-complete reservoir of gas, but form a massive, diffuse clump.  If we exclude it, then a low-mass clump will be extracted, but it will not represent a complete reservoir, and there is even no guarantee that it will be representative of the star-forming gas.

\figref{fig:masshistoINC} shows a histogram of the augmented clump masses, extracted without the 90 per cent threshold.  
\begin{figure} 
\centering
\includegraphics[width=\columnwidth]{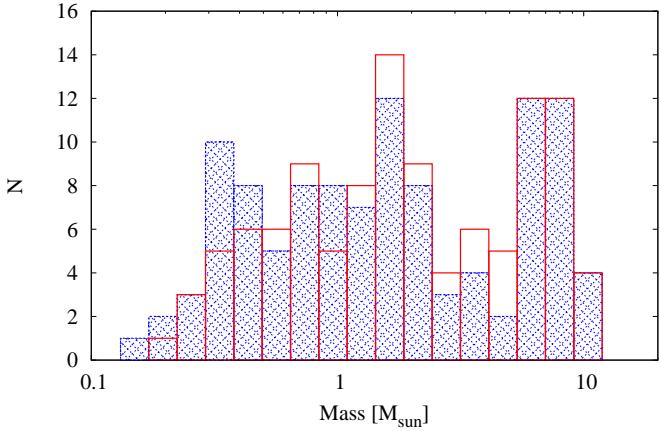}
\caption{Histogram of the clump masses, where we have excluded the requirement that 90 per cent of the associated gas must be in the final clump (blue hatched).  For reference, we have included the mass distribution of the fully augmented clumps as in \figref{fig:masshisto}.  Comparing the two distributions shows that removing the threshold removes several diffuse clumps (notably from \hyd{} and \ntwenty{}) and increases the number of low-mass clumps. }
\label{fig:masshistoINC}
\end{figure} 
Clearly, the massive clumps ($M \gtrsim 5$~\Msun{}) are unaffected by this criteria. However, the diffuse clumps ($2 \lesssim M$/\Msun{}$\lesssim 5$) are removed by excluding this criteria; there remain a few clumps in this mass range (mainly from \itwenty{}), but they are not diffuse.  The majority of these diffuse clumps that have been removed are from \hyd{} and \ntwenty{}.   

These former diffuse clumps (i.e., the low-mass clumps formed from the progenitors of diffuse clumps but with the 90 per cent threshold excluded) do not satisfy the trends as discussed in \secref{sec:results}.  Indeed, they become even more notable outliers in many of our trends.  They are still extracted at an early time and there is still \sm80~kyr between extraction time and the formation of their progenitor sink.  They are larger than other clumps of similar mass, suggesting that our neighbour search is finding the associated gas that exists in the filaments; therefore, much of this gas still comes in from far away, albeit not as far as in the full diffuse clump.  As discussed above with the non-diffuse clumps, gas is continually accreting on to the filaments, thus extracting a filamentary structure alone represents an incomplete reservoir.  Moreover, since this extracted gas is filamentary, it has higher velocity dispersions, rms Mach numbers, and ratios $E_\text{kin}/\left|E_\text{pot}\right|$ and $E_\text{mag}/\left|E_\text{pot}\right|$ than other clumps of similar masses.   

Therefore, with these progenitors, we have the choice of creating diffuse clumps that include a near-complete reservoir but may better represent a different progenitor star, or lower-mass but incomplete reservoirs that do not follow the general trends of the remaining clumps.  The former option yields unrepresentative clumps that are too massive to reasonable re-simulate.  The latter option yields clumps with larger-than-expect sizes and long time between extraction and sink formation, which suggests that these clumps do not well-represent the star-forming gas of their progenitors.  Therefore, we conclude that both diffuse and former diffuse clumps are not representative of their progenitor's star forming gas, and should be excluded in future studies and analyses.

\section{Summary and conclusion}
\label{sec:conc}

In this paper, we created initial star forming clumps by selecting and extracting gaseous regions from the low-mass star cluster simulations of \citet{WursterBatePrice2019}; that numerical study employed the smoothed particle magnetohydrodynamics method.  For each star (i.e., sink particle) that formed in each of the nine simulations, we identified all the gas particles that either accreted onto the star or were bound to the star; we did this for the entire lifetime of the star.  We then tracked this gas backwards in time to before the star formed and to when the maximum density of this gas dropped below \rhoxls{-16}.  We finally augmented this gas with the neighbouring gas to obtain a true representation of the star forming clump.  From \citetalias{WursterBatePrice2019}, we extracted 109 clumps that spanned a mass range of $0.15 \lesssim M/$\Msun{}$ \lesssim 10.2$; 19 clumps were from a purely hydrodynamical simulation thus had no magnetic properties.   

In this paper, we analysed these resulting clumps at the time they were extracted from the parent simulation, while in \citetalias{WursterRowan2023b} we will re-simulate several of these clumps to compare the evolution of the clump in isolation compared to that in the original cluster environment.  Our main conclusions are as follows:  

\begin{enumerate}
\item More massive clumps are more spherical while smaller clumps are more triaxial.  When massive clumps are extracted at early times, the gas is approximately isotropically located within the parent cloud.  When the smaller clumps are extracted at later times, much, but not all, of the gas that will interact with the progenitor star is in the filaments.  Therefore, well-defined filaments alone do not contain the entire reservoir of gas for a star.

\item The gas density in the clumps typically spans a range of 2--4 orders of magnitude, with 95.4 per cent of the gas spanning $\gtrsim2$ orders of magnitude; this range is much larger than that of a typical Bonnor-Ebert sphere or (obviously) a sphere of uniform density.  There is no common density distribution amongst the clumps.  The gas distributions are not smooth, demonstrating that the clumps are highly structured at extraction time, although they tend to be smoother for smaller clumps.

\item The extracted star forming clumps are turbulent with velocity dispersions that follows Larson's law.  The rms Mach number of the clump increases with clump mass, and is generally lower than the Mach number of the parent cloud at extraction time for the lower-mass clumps.  At extraction time, the clumps are purely turbulent, with no coherent rotation.  

\item All clumps have a similar range of magnetic field strengths.  There is a slight trend that clumps extracted from parent clouds with stronger initial magnetic field strengths themselves have stronger magnetic fields; however, this difference is smaller than the initial difference amongst the parent clouds, indicating that the clumps lose some memory of their larger environment.   There is no consistent distribution of magnetic field strengths, although the distributions are generally smooth, uni-modal, and span an order of magnitude.  This suggest that there is some coherent ordering of the magnetic field at extraction time.

\item The clumps contain very little memory of the initial magnetic field geometry; there is a broad distribution of angles between the extracted magnetic field and the parent's initial magnetic field within each clump, where the distributions cover all possible angles.  There is generally no well-defined peak in the distribution.  Although the field may be ordered, is it not reflective of the initial field order.

\end{enumerate}

Throughout our analysis, we find that the properties of the extracted clumps are independent (or only weakly dependent) on the initial magnetic field strength of the parent cloud; moreover, the properties are independent of whether or not the parent cloud was evolved using non-ideal MHD.  Most properties have a slight trend with clump mass, which is reasonable given that the more massive clumps contain a wider distribution of gas and considerable quantities of low-density gas.  This low-density gas both broadens and skews the distributions and modifies the average values compared to small, compact clumps.  Given that any trend is weak, we conclude that stars are born from a wide variety of environments and there is not a single universal star forming clump. 

\section*{Acknowledgements}

We would like to thank the anonymous referee for useful comments that improved the quality of this manuscript.
JW acknowledges support from the University of St Andrews.
CR acknowledges support from the European Research Council (ERC) under the European Union’s Horizon 2020 research and innovation program under grant agreement No 638435 (GalNUC).
This work was performed using the DiRAC Data Intensive service at Leicester, operated by the University of Leicester IT Services, which forms part of the STFC DiRAC HPC Facility (www.dirac.ac.uk). The equipment was funded by BEIS capital funding via STFC capital grants ST/K000373/1 and ST/R002363/1 and STFC DiRAC Operations grant ST/R001014/1. DiRAC is part of the National e-Infrastructure.
In order to meet institutional and research funder open access requirements, any accepted manuscript arising shall be open access under a Creative Commons Attribution (CC BY) reuse licence with zero embargo.  
Several figures were made using \textsc{splash} \citep{Price2007}.  

\section*{Data availability}

The data underlying this article and \citet{\wbp2019} will be available upon reasonable request.
\bibliography{ResimulationOne.bib}
\appendix
\section{All clumps}
\figref{fig:morph:all} shows all 109 clumps extracted from all nine simulations in \citetalias{WursterBatePrice2019}.  From each parent cloud, there are at least a few very similar clumps due to stellar multiplicity.   These clumps generally contain a high fraction of mutual gas, and the more massive clumps often contain nearly all the gas of similar, less massive clumps.  Despite their similarities, however, no two clumps are identical, even if they have the same extraction time.

\begin{figure*} 
\centering
\includegraphics[width=\textwidth]{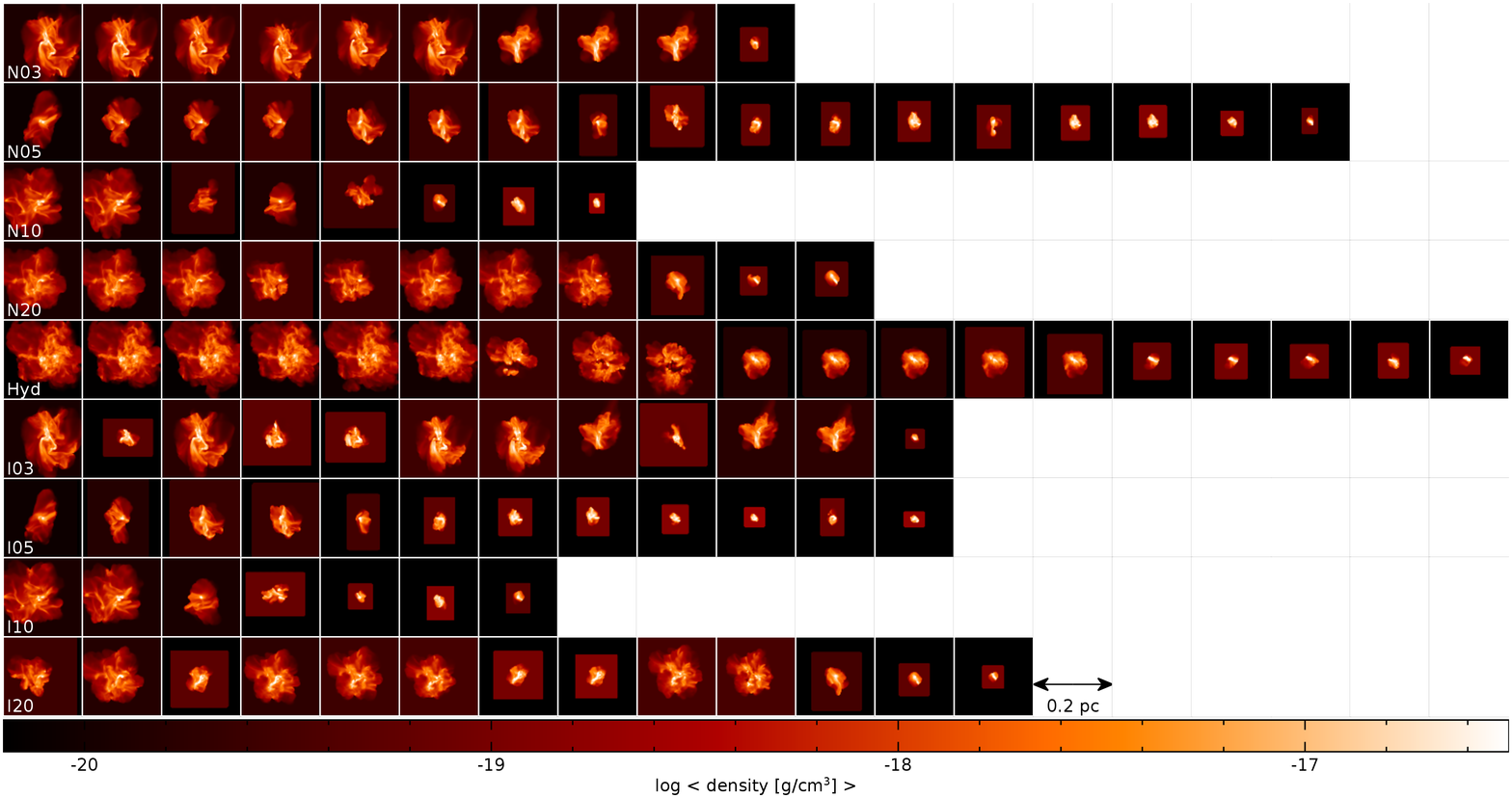}
\includegraphics[width=\textwidth]{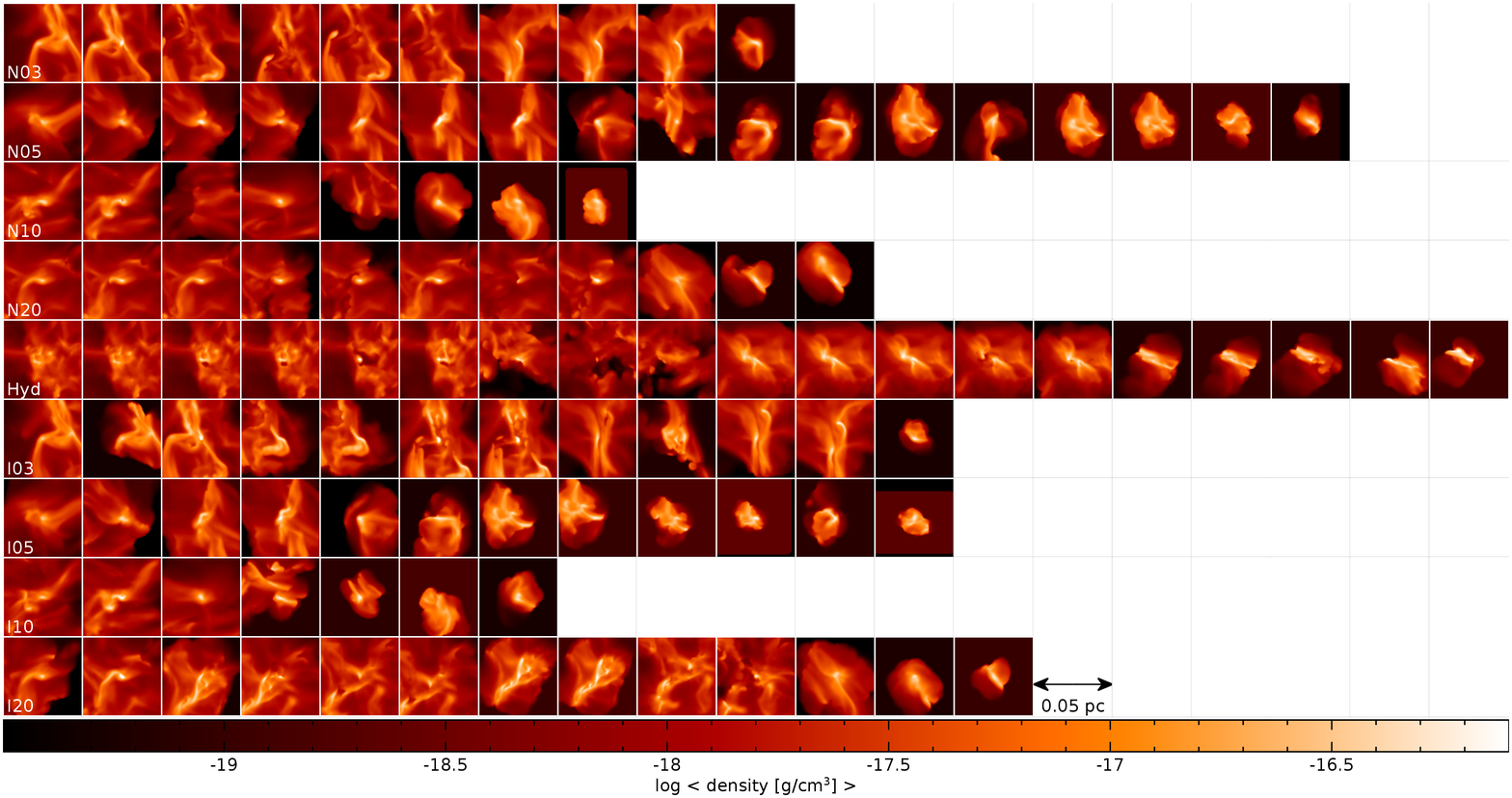}
\caption{Average gas density in the $xz$-plane of all  the augmented clumps.   From left to right, clumps are ordered from earliest to latest extraction time.  The clumps are shown at two different scales to highlight the large (top) and small (bottom) clumps.  From a given parent cloud, there are many similar clumps due to the hierarchal nature of the stellar systems at $t_\text{final}$.}
\label{fig:morph:all}
\end{figure*} 
\label{lastpage}
\end{document}